# Computing with Harmonic Functions

## Sheldon Axler

*1 November 2016*          **© Sheldon Axler**

---

## Getting Started

---

## Harmonic Function Theory

This document is the manual for HFT11.m (version 11.00), a *Mathematica* package for computing with harmonic functions. The HFT11.m package allows the user to make calculations that would take a prohibitive amount of time if done without a computer. For example, the Poisson integral of any polynomial can be computed exactly.

Earlier versions of this software focused on algorithms arising from the material in the book *Harmonic Function Theory* [ABR] by Sheldon Axler, Paul Bourdon, and Wade Ramey. That book is still the source for many of the algorithms used in the HFT11.m package, but the goal of the package has expanded to include additional symbolic manipulations involving harmonic functions. The HFT11.m package can be used without the *Harmonic Function Theory* book, but the debt to the book is reflected in the initials chosen for the name of the package.

This document, the HFT11.m package, and the HFT11.nb notebook are intended for use with *Mathematica* versions 7 through 11 (the package will probably also work with later versions of *Mathematica*).

**This document (which is available in both nb and pdf formats) and the HFT11.m package and the HFT11.nb notebook that serves as the source file for the HFT11.m package are available electronically free of charge. The most recent versions are available at http : // www.axler.net / HFT_Math.html. New features are added**

## periodically, so check for new versions occasionally.

Comments, suggestions, and bug reports about this document, the HFT11.m package, or the book *Harmonic Function Theory* are welcome; please send them to axler@sfsu.edu.

Work on some of the algorithms used in this package was partially supported by research grants to the author from the National Science Foundation.

## Loading the Package

To begin a session with the HFT11.m package, first start *Mathematica*. Then enter the command shown below, making an appropriate modification to indicate the directory or folder in which HFT11.m is located on your computer. Or, after typing <<, you can use the File Path command, located on the Insert menu, to paste the full pathname of HFT11.m into your input cell:

*In[1]:=* `<< "C:\\Box Sync\\math\\publications\\HFT-Mathematica\\HFT11.m"`

HFT11.m, version 11.00, 1 November 2016; for use with *Mathematica*
  versions 7 through 11 (and probably later versions of *Mathematica*).

The HFT11.m *Mathematica* package is designed for computing with harmonic functions.

Documentation for the use of this package and information about the
  algorithms used in it is available in the document titled *Computing*
  *with Harmonic Functions*, which is available in both nb and pdf formats.

The most recent version of this HFT11.m package and its documentation
  *Computing with Harmonic Functions* are available at http://www.axler.net.

For addional information about harmonic functions, see
  the book *Harmonic Function Theory* (second edition), by Sheldon
  Axler, Paul Bourdon, and Wade Ramey, published by Springer.

This package is copyrighted by Sheldon Axler but is distributed without charge.

Comments, suggestions, and bug reports should be sent to axler@sfsu.edu.

* You can now use the functions in this package.

After loading the HFT11.m package, you can use the functions described in the rest of this document.

The nb version of this document is a live *Mathematica* notebook, meaning that you can and should evaluate and modify the input cells in this notebook. The pdf version of this document is a static document. The nb and pdf versions of this document and the HFT package are revised periodically with the addition of new features; new versions are posted at http : // www.axler.net/HFT_Math.html.

## On-line Help

This document provides the only complete documentation for using the HFT11.m package. However, users can be reminded about the basic use and syntax of each function created by HFT11.m by entering ? followed by the function name:

*In[2]:=*
```
? volume
```

volume[n] gives the volume of the unit ball in n−dimensional real Euclidean space.

The *Mathematica* statement Names["HFT`*"] will produce a complete list of the functions created by HFT11.m:

*In[3]:=*
```
Names["HFT`*"]
```

*Out[3]=*
{annulus, antiLaplacian, basisH, bergmanKernel, bergmanKernelH,
bergmanProjection, biDirichlet, dimension, dimHarmonic, dirichlet, divergence,
expandNorm, exteriorNeumann, exteriorSphere, gradient, harmonicConjugate,
harmonicDecomposition, hilbertSchmidt, homogeneous, integrateBall,
integrateEllipsoidArea, integrateEllipsoidVolume, integrateSphere, jacobian,
kelvin, kelvinH, laplacian, multiple, neumann, norm, normalD, partial,
poissonKernel, poissonKernelH, quadratic, reflection, region, schwarz,
setDimension, singularity, southPole, surfaceArea, taylor, togetherness,
trace, turnOff, turnOn, volume, zeroToZero, zonalHarmonic, $\delta$, $\triangle$, $\Phi$}

The naming scheme for functions introduced by this package is that function names are generally spelled out in full and begin with a lower-case letter, as in **laplacian**. This is a major change from HFT9 and previous versions of the package, where function names began with an upper-case letter. This change was made because new versions of *Mathematica* have introduced new functions beginning with an upper-case letter, which have clashed with names already used by the HFT package. With this change (meaning that HFT functions now begin with a lower-case letter), this issue has disappeared.

Functions in the HFT11 package with a name formed from more than one word begin with a lower-case letter but then have the first letter of additional words begin with an upper-case letter, as in **surfaceArea**.

# Calculus in R$^n$

## norm

**`norm[x]`** is the Euclidean norm of a vector *x*. When using StandardForm for output (which is the default in *Mathematica*), this package will use **‖x‖** instead of **`norm[x]`**. When using StandardForm for input (which is the default in *Mathematica*), you can use either **‖x‖** or **`norm[x]`** :

*In[4]:=*  ‖**5 x**‖ **− norm[2 x]²**

*Out[4]=*  5 ‖x‖ − 4 ‖x‖²

To type **‖x‖**, type escape-key **`l | |`** escape-key (the three entries between the two escape-keys are one lower-case el followed by two |), then **x**, then escape-key **`r | |`** escape-key (the three entries between the two escape-keys are one lower-case r followed by two |). To help you remember, el is associated with the left norm bracket and r is associated with the right norm bracket. Even though **‖x‖** looks better, it may be easier to type **`norm[x]`**.

## expandNorm

**expandNorm** expands norms of sums (and differences) by replacing expressions of the form $\|x+y\|$ with $(\|x\|^2 + \|y\|^2 + x.y)^{1/2}$. **expandNorm** is often useful for verifying identities. For example, to prove the Symmetry Lemma (1.13 in [ABR]), which states that

$$\left\| \frac{y}{\|y\|} - \|y\| \, x \right\| = \left\| \frac{x}{\|x\|} - \|x\| \, y \right\|$$

for all nonzero $x, y \in \mathbf{R}^n$, we subtract the left-hand side of the alleged equality from the right-hand side, hoping that the result equals 0:

*In[5]:=*  $\left\| \dfrac{x}{\|x\|} - \|x\| \, y \right\| - \left\| \dfrac{y}{\|y\|} - \|y\| \, x \right\|$

*Out[5]=*  $\left\| \dfrac{x}{\|x\|} - y \, \|x\| \right\| - \left\| \dfrac{y}{\|y\|} - x \, \|y\| \right\|$

The last output does not look like 0, so we apply **expandNorm** to it. The symbol **%** used below is the *Mathematica* abbreviation for the last output:

| | |
|---|---|
| *In[6]:=* | `expandNorm[%]` |
| *Out[6]=* | `0` |

Because the last output equals 0, the Symmetry Lemma has been proved.

---

## **partial**

**partial[f, x_j]** is the partial derivative of $f$ with respect to $x_j$. Here $x$ denotes a vector in Euclidean space, and $x_j$ is the $j^{\text{th}}$ coordinate of $x$. The advantage of using **partial** rather than the built-in Mathematica function **D** or **∂** is that **partial** knows that

$x_j$ is the $j^{\text{th}}$ coordinate of $x$, and furthermore **partial** can deal with norms and dot products. Note that the dot product of $b$ and $x$ is denoted by **b.x**:

| | |
|---|---|
| *In[7]:=* | `partial[‖a + x‖^{b.x}, x_5]` |
| *Out[7]=* | `Log[‖a + x‖] ‖a + x‖^{b.x} b_5 + b.x ‖a + x‖^{-2+b.x} (a_5 + x_5)` |

A subscript can be typed in *Mathematica* by pressing control-dash. Return to normal mode by pressing control-space. A superscript or exponent can be typed by pressing control-6. Instead of entering **x_j**, you can type **x[j]**.

For taking multiple partial derivatives, **partial** uses the same syntax as the built-in *Mathematica* function **D**. For example, the following command computes the partial derivative of $\|x\|$ once with respect to the second coordinate, twice with respect to the first coordinate, and three times with respect to the fourth coordinate:

| | |
|---|---|
| *In[8]:=* | `partial[‖x‖, x_2, {x_1, 2}, {x_4, 3}]` |
| *Out[8]=* | $-\dfrac{1}{‖x‖^{11}}15 \left(3 ‖x‖^4 x_2 x_4 - 21 ‖x‖^2 x_1^2 x_2 x_4 - 7 ‖x‖^2 x_2 x_4^3 + 63 x_1^2 x_2 x_4^3\right)$ |

## Δ (`laplacian`)

$\Delta_x$ [**f**] is the Laplacian of $f$ with respect to $x$:

*In[9]:=*
$$\Delta_x \, [\, x_3 \, \|x\| \, ]$$

*Out[9]=*
$$\frac{x_3 + \text{dimension}[x] \, x_3}{\|x\|}$$

To type $\Delta$, type escape-key D escape-key. Instead of $\Delta$, you can use **laplacian**; thus **laplacian**$_x$[**f**] is the same as $\Delta_x$[**f**].

Note that in the last result, the Laplacian depends upon `dimension[x]`, which is the dimension of the Euclidean space in which $x$ lives. For work in $\mathbf{R}^n$, **setDimension[x, n]** should be used to set the dimension of $x$ equal to $n$; here $n$ can be a symbol or a specific integer, say 8. Be sure to use the **setDimension** function rather than changing the value of `dimension[x]` directly. We illustrate this procedure by showing that for each $\zeta$ with $|\zeta| = 1$, the Poisson kernel $P(\,.\,, \zeta)$ is harmonic (in other words, what follows is a proof of Proposition 1.18 of [ABR]):

*In[10]:=*
**setDimension[x, n]**

··· setDimension:
  x will be considered to be a vector in n–dimensional real Euclidean space.

*In[11]:=*
$$\Delta_x \, \big[ \, \big( 1 - \|x\|^2 \big) \, \big/ \, \|x - \varsigma\|^n \, \big]$$

*Out[11]=*
$$4 \, n \, \big( -x \, . \, \varsigma + \|x\|^2 \big) \, \|x - \varsigma\|^{-2-n} - 2 \, n \, \|x - \varsigma\|^{-n} + \big( 1 - \|x\|^2 \big) \, \big( -(-1 + n) \, n \, \|x - \varsigma\|^{-2-n} - n \, (1 + n) \, \|x - \varsigma\|^{-4-n} \big( 2 \, x \, . \, \varsigma - \|x\|^2 - \|\varsigma\|^2 \big) \big)$$

*In[12]:=*
**expandNorm[%] /. $\|\varsigma\| \to 1$**

*Out[12]=*
0

Thus the Laplacian of $P(\,.\,, \zeta)$ is 0, and hence $P(\,.\,, \zeta)$ is harmonic, as claimed.

If $x$ and $y$ are symbols, then $\Delta_{x,y}[f]$ gives the Laplacian of $f$ with respect to the vector $(x, y)$, where $x$ is thought of as a vector, $y$ is thought of as a real variable, and $(x, y)$ is thought of as a vector in a Euclidean space whose dimension is one more than the dimension of $x$. This format is often useful when working with functions defined on the upper half-space, as in Chapter 7 of [ABR]. Note that if we are thinking of $(x, y)$ as an element of $\mathbf{R}^n$, then $x$ should be defined to have dimension $n - 1$. We illustrate the use of this format of the $\Delta$ function by showing that for each $t \in \mathbf{R}^{n-1}$, the Poisson kernel $P_H(\ .\ ,\ t)$ for the upper half-space is harmonic (see page 145 of [ABR] for another proof of this):

*In[13]:=* 
```
setDimension[x, n - 1]
```

> ••• setDimension:
> x will be considered to be a vector in −1 + n−dimensional real Euclidean space.

*In[14]:=* 
```
Δx,y[y / (‖x − t‖² + y²)^(n/2)]
```

*Out[14]=* 
$$-\frac{1}{\|-t + x\|^2}$$
$$n\,y\,\left(y^2 + \|-t + x\|^2\right)^{-2-\frac{n}{2}}\,\left(2\,t.x - \|t\|^2 - \|x\|^2 + \|-t + x\|^2\right)\,\left(-y^2 + \|-t + x\|^2 + n\,\|-t + x\|^2\right)$$

*In[15]:=* 
```
expandNorm[%]
```

*Out[15]=* 
```
0
```

Thus the Laplacian of $P_H(\ .\ ,\ t)$ is 0, and hence $P_H(\ .\ ,\ t)$ is harmonic, as claimed.

The expression $\Delta_v[f]$, where $v$ is a list of explicit coordinates, gives the Laplacian of $f$ with respect to the coordinates in the list $v$. For example, below we compute the Laplacian of $x^2\,y^3\,z^4$ with respect to the usual coordinates $x$, $y$, and $z$ of $\mathbf{R}^3$:

*In[16]:=* 
```
Δ{x,y,z}[x² y³ z⁴]
```

*Out[16]=* 
$$12\,x^2\,y^3\,z^2 + 6\,x^2\,y\,z^4 + 2\,y^3\,z^4$$

Note that in the example above, $x$ is no longer being used as a vector but as the first coordinate in $\mathbf{R}^3$. Because $x$ has been used in a list of coordinates, the software no longer treats $x$ as a vector.

If $m$ is a positive integer, then $\Delta_x{}^\mathbf{m}$ can be used, with the same syntax as $\Delta_x$, to evaluate the $m^{\text{th}}$ power of the Laplacian. For example, here we find the biLaplacian (the Laplacian of the Laplacian) of $1/|x|$ on $\mathbf{R}^8$:

*In[17]:=*    `setDimension[x, 8]`

   ··· setDimension:

     x will be considered to be a vector in 8–dimensional real Euclidean space.

*In[18]:=*    $\Delta_x{}^2[1 / \|x\|]$

*Out[18]=*    $\dfrac{45}{\|x\|^5}$

## **gradient**

**gradient[f, x]** is the gradient of $f$ with respect to the vector $x$:

*In[19]:=*    `gradient[a.x + x_k + ‖x‖, x]`

*Out[19]=*    $a + \dfrac{x}{\|x\|} + \delta_k$

In the last output, $\delta_k$ denotes the vector that equals 1 in the $k^{\text{th}}$ coordinate and 0 in the other coordinates. To type $\delta$, type escape-key d escape-key.

## **normalD**

**normalD[f, z]** equals the outward normal derivative of $f$ as a function of $z$, where we also think of $z$ as a point on the unit sphere in a Euclidean space.

In the section of Chapter 1 of [ABR] titled *The Poisson Kernel for the Ball*, the formula for the Poisson kernel was derived by showing that $P(x, z)$ equals the normal derivative (with respect to $z$) of

$$\left( \|z - x\|^{2-n} - \|x\|^{2-n} \left\| z - \frac{x}{\|x\|^2} \right\|^{2-n} \right) \Big/ (2 - n).$$

The computation of that normal derivative is a bit complicated and was left to the reader. Now we show how this computation can be done easily by a computer. We begin by taking the normal derivative of the function above:

*In[20]:=*
```
normalD[ (∥z - x∥^(2-n) - ∥x∥^(2-n) ∥z - x/∥x∥^2 ∥^(2-n)) / (2 - n) , z]
```

*Out[20]=*
$$\frac{1}{2-n} \left( (2-n) \ (1-x.z) \ \|-x+z\|^{-n} - (-2+n) \ \|x\|^{-n} \ (x.z - \|x\|^2) \ \left\| z - \frac{x}{\|x\|^2} \right\|^{-n} \right)$$

Now we ask the computer to replace $\left\| z - \frac{x}{\|x\|^2} \right\|$ in the last output with $\|z - x\| / \|x\|$; the Symmetry Lemma (1.13 of [ABR]) implies that the two expressions are equal:

*In[21]:=*
```
% /. ∥z - x/∥x∥^2 ∥ → ∥z - x∥ / ∥x∥
```

*Out[21]=*
$$\frac{1}{2-n} \left( (2-n) \ (1-x.z) \ \|-x+z\|^{-n} + (-2+n) \ \|x\|^{-n} \ (-x.z + \|x\|^2) \ \left( \frac{\|-x+z\|}{\|x\|} \right)^{-n} \right)$$

Applying the *Mathematica* function `PowerExpand` to the last output gives the desired result:

*In[22]:=*
```
PowerExpand[%]
```

*Out[22]=*
$$- \left( -1 + \|x\|^2 \right) \ \|-x+z\|^{-n}$$

The last output completes the derivation of the formula for the Poisson kernel for the ball.

`normalD[f, q, z]` equals the normal derivative of $f$ (as a function of $z$) with respect the surface $q(z) = c$ for some constant $c$. In other words, `normalD[f, q, z]` equals $(\nabla f . \nabla q) / \|\nabla q\|$.

For example, this expression computes the normal derivative of $x_1^2 \, x_2^8 \, x_3^5$ on the ellipsoid $x_1^2 + 3\,x_2^2 + 2\,x_3^2 = 1$:

*In[23]:=*
```
normalD[x_1^4 x_2^8 x_3^5, x_1^2 + 3 x_2^2 + 2 x_3^2, x]
```

*Out[23]=*
$$\frac{38 \, x_1^4 \, x_2^8 \, x_3^5}{\sqrt{x_1^2 + 9 \, x_2^2 + 4 \, x_3^2}}$$

## Matrices

### Matrix dimensions

*Mathematica* uses `.` to represent matrix multiplication as well as dot product. When using this package, you must use the **`setDimension`** command to tell *Mathematica* whenever a symbol is being thought of as a matrix. To think of *A* as an *m*-by-*n* matrix, enter the command **`setDimension[A, {m, n}]`**; here *m* and *n* can be symbols or specific positive integers such as 7 and 3. For the examples in this section, we want to think of *A* as a square matrix, so we make the two dimensions of *A* equal:

*In[24]:=*  `setDimension[A, {n, n}]`

⋯ setDimension: A will be considered to be a matrix of size n–by–n.

*In[25]:=*  `gradient[‖A.x‖, x]`

*Out[25]=*
```
A.x.A
―――――
‖A.x‖
```

To interpret input or output involving products of matrices and vectors, think of vectors as either column vectors or row vectors, whichever makes sense in context. Thus in the input above, *x* is obviously a vector, because we are taking the gradient of some function with respect to *x*. For *A.x* to make sense as a vector in $\mathbf{R}^n$ in the input above, we must think of *x* as a column vector (an *n*-by-1 matrix). To interpret (*A.x*).*A* in the output above, we again think of *x* as a column vector (an *n*-by-1 matrix), so that *A.x* makes sense as an element of $\mathbf{R}^n$. Because *A.x* is then multiplied by the *n*-by-*n* matrix *A*, we must think of *A.x* as a row vector (an 1-by-*n* matrix), so that (*A.x*).*A* is a vector, as expected here (because it is the gradient of a function).

In the example above we did not enter the command **`setDimension[x, n]`** to tell *Mathematica* that *x* is a vector (although it would not have hurt to do so). The HFT11.m package can almost always use the context to distinguish scalars from vectors (if both interpretations make sense, the package assumes that a symbol represents a scalar). However, you must always explicitly use the **`setDimension`** command to tell *Mathematica* which objects are matrices.

The expression `dimension[A]` gives the dimensions of a matrix $A$. However, be sure to use the `setDimension` function rather than changing the value of `dimension[A]` directly.

### Rows and columns

If $A$ is a matrix, then $A_{j,\cdot}$ denotes the $j^{\text{th}}$ row of $A$ and $A_{\cdot,j}$ denotes the $j^{\text{th}}$ column of $A$:

*In[26]:=*
```
partial[‖A.x‖ + 5 ‖x.A‖, x₃]
```

*Out[26]=*
$$\frac{A.x.A_{\cdot,3}}{\|A.x\|} + \frac{5\, x.A.A_{3,\cdot}}{\|x.A\|}$$

Note that in the input above, $x$ is thought of as a column vector in the term $A.x$ and as a row vector in the term $x.A$. To interpret $(A.x).A_{\cdot,3}$ in the output above, read from left to right and think of $x$ as a column vector, so that $A.x$ is a vector; then $(A.x).A_{\cdot,3}$ is just the dot product of two vectors, so that it is a number, as expected here. Similarly, to interpret $(x.A).A_{\cdot,3}$ in the output above, read from left to right and think of $x$ as a row vector, so that $x.A$ is a vector; then $(x.A).A_{3,\cdot}$ is just the dot product of two vectors, so that it is a number, as expected here

### Hilbert Schmidt norm

`hilbertSchmidt[A]` is the Hilbert-Schmidt norm of a matrix $A$, which is the square root of the sum of the squares of all entries of $A$:

*In[27]:=*
```
hilbertSchmidt[4 IdentityMatrix[n] ]
```

*Out[27]=*
```
4 n
```

*In[28]:=*
```
Δₓ[‖x.A‖]
```

*Out[28]=*
$$\frac{\text{hilbertSchmidt}[A]^2 \, \|x.A\|^2 - \|x.A.\text{Transpose}[A]\|^2}{\|x.A\|^3}$$

### Trace

`trace[A]` is the trace of a square matrix $A$:

In[29]:= `trace[-5 IdentityMatrix[n] ]`

Out[29]= $-5\,n$

---

## **divergence**

**`divergence[f, x]`** is the divergence of *f* with respect to the vector *x*:

In[30]:= `setDimension[A, {n, n}]`

··· setDimension: A will be considered to be a matrix of size n–by–n.

In[31]:= `divergence[A.x / ‖x‖, x]`

Out[31]= $\left(-\dfrac{x}{\|x\|^3}\right).A.x + \dfrac{\texttt{trace}[A]}{\|x\|}$

---

## **jacobian**

**`jacobian[f, x]`** is the Jacobian derivative of *f* with respect to *x*. Here *f* should be a function of *x* taking values in some Euclidean space, and the Jacobian derivative is the usual matrix consisting of partial derivatives of the coordinates of *f*:

In[32]:= `setDimension[x, n]`

··· setDimension:
  x will be considered to be a vector in n–dimensional real Euclidean space.

In[33]:= `jacobian[x / ‖x‖², x]`

Out[33]= $-\dfrac{2\,\texttt{Transpose}[\{x\}].\{x\}}{\|x\|^4} + \dfrac{\texttt{IdentityMatrix}[n]}{\|x\|^2}$

Although $x$ can be thought of as a row vector or a column vector, depending on the context, if we were using explicit coordinates we would represent $x$ in *Mathematica* by something like `{x₁, x₂, ..., xₙ}`. Thus `{x}` would be `{{x₁, x₂, ..., xₙ}}`, which is a 1-by-$n$ matrix. Thus `Transpose[{x}]` is an $n$-by-1 matrix, and the product `Transpose[{x}].{x}`, which appears in the last output, is an $n$-by-$n$ matrix, as expected.

## homogeneous

**homogeneous[u, m, x]** is the term of degree $m$ in the homogeneous expansion of $u$ at the origin; here $u$ is thought of as a function of the vector $x$:

*In[34]:=* 
```
setDimension[x, 3]
```

⋯ **setDimension**:
  x will be considered to be a vector in 3−dimensional real Euclidean space.

*In[35]:=* 
```
homogeneous[(1 - ||x||²) / Cos[x₁]^(3/2), 4, x]
```

*Out[35]=* $\frac{1}{32} \left( -11 x_1^4 - 24 x_1^2 x_2^2 - 24 x_1^2 x_3^2 \right)$

To find homogeneous expansions about points other than the origin, use **Homogeneous[u, m, x, b]**, which gives the term of degree $m$ in the homogeneous expansion of $u$, as a function of $x$, about $b$:

*In[36]:=* 
```
homogeneous[x₁ x₂ + x₃⁴, 2, x, b]
```

⋯ **togetherness**: togetherness has been turned
  off.  The command  turnOn[togetherness]  will turn it back on.

*Out[36]=* $\left( -b_1 + x_1 \right) \left( -b_2 + x_2 \right) + 6 b_3^2 \left( -b_3 + x_3 \right)^2$

The first time in each *Mathematica* session that you use **homogeneous** to find an expansion about a point other than the origin, you may see a message that `togetherness` has been turned off.  The togetherness subsection later in this document provides an explanation for this message.

## **taylor**

**taylor[u, m, x]** is the sum of all terms of degree at most *m* in the Taylor series expansion of *u* at the origin; here *u* is thought of as a function of the vector *x*:

*In[37]:=* 
```
taylor[x₁²/(6 + x₂ + Cos[x₃]), 4, x]
```

*Out[37]=* 
$$\frac{x_1^2}{7} - \frac{1}{49} x_1^2 x_2 + \frac{1}{343} x_1^2 x_2^2 + \frac{1}{98} x_1^2 x_3^2$$

To find Taylor series expansions about points other than the origin, use **taylor[u, m, x, b]**, which gives the sum of all terms of degree at most *m* in the Taylor series expansion of *u*, as a function of *x*, about *b*:

*In[38]:=* 
```
taylor[1 + x₁ x₂ + x₁², 2, x, b]
```

*Out[38]=* 
$$1 + b_1^2 + b_1 b_2 + (2 b_1 + b_2) (-b_1 + x_1) + (-b_1 + x_1)^2 + b_1 (-b_2 + x_2) + (-b_1 + x_1) (-b_2 + x_2)$$

The first time in each *Mathematica* session that you use **Taylor** to find an expansion about a point other than the origin, you may see a message that togetherness has been turned off. The togetherness subsection later in this document provides an explanation for this message. For now, we turn togetherness back on so that further output in this document will be displayed nicely.

*In[39]:=* 
```
turnOn[togetherness]
```

••• **togetherness**: togetherness has been turned on. The command turnOff[togetherness] will turn it back off.

## **Nonexplicit functions**

Our examples so far have involved only concretely defined functions. But functions in symbolic form can also be used with each of the differentiation commands in this package. Here, for example, is how to find the gradient of the function whose value at *x* is $f\big((g(3\,x))^2\big)$:

*In[40]:=*
```
gradient[f[g[3 x]²], x]
```

*Out[40]=*
```
6 g[3 x] gradient[g][3 x] f′[g[3 x]²]
```

Here `gradient[g][3x]` denotes the gradient of $g$, evaluated at $3\,x$. For the example above to make sense, we (and the computer) must think of $g$ as a real-valued function on some Euclidean space (where $x$ lives) and $f$ as a function from $\mathbf{R}$ to $\mathbf{R}$.  Let's find the Laplacian of the same function:

*In[41]:=*
```
Δₓ[f[g[3 x]²]]
```

*Out[41]=*
```
18 (∥gradient[g][3 x]∥² + g[3 x] Δ[g][3 x]) f′[g[3 x]²] +
  36 g[3 x]² ∥gradient[g][3 x]∥² f″[g[3 x]²]
```

`Δ[g][3x]` denotes, of course, the Laplacian of $g$, evaluated at $3\,x$.

All our differentiation commands (**partial**, **Δ**, **gradient**, **normalD**, **divergence**, **Jacobian**) and **homogeneous** and **taylor** can be used with non-explicit functions.  For example, if working in $\mathbf{R}^2$, to find the term of degree 3 in the homogeneous expansion of $f\big((g(3\,x))^2\big)$ about the origin, enter the commands shown below. In the output below you will see terms such as `partial_{1,1,2}[g][0]`, which denotes the partial derivative of $g$, twice with respect to the first variable and once with respect to the second variable, evaluated at 0:

*In[42]:=*
```
setDimension[x, 2]
```

⋯ setDimension:

    x will be considered to be a vector in 2–dimensional real Euclidean space.

*In[43]:=* `homogeneous[ f[g[3 x]²], 3, x]`

*Out[43]=* $9 x_1^3 \left( 6 g[0] f''[g[0]^2] \text{ partial}_{(1)}[g][0]^3 + 4 g[0]^3 f^{(3)}[g[0]^2] \text{ partial}_{(1)}[g][0]^3 + \right.$
$3 f'[g[0]^2] \text{ partial}_{(1)}[g][0] \text{ partial}_{(1,1)}[g][0] + 6 g[0]^2 f''[g[0]^2]$
$\text{partial}_{(1)}[g][0] \text{ partial}_{(1,1)}[g][0] + g[0] f'[g[0]^2] \text{ partial}_{(1,1,1)}[g][0] \left. \right) +$
$27 x_1^2 x_2 \left( 6 g[0] f''[g[0]^2] \text{ partial}_{(1)}[g][0]^2 \text{ partial}_{(2)}[g][0] + \right.$
$4 g[0]^3 f^{(3)}[g[0]^2] \text{ partial}_{(1)}[g][0]^2 \text{ partial}_{(2)}[g][0] +$
$f'[g[0]^2] \text{ partial}_{(2)}[g][0] \text{ partial}_{(1,1)}[g][0] +$
$2 g[0]^2 f''[g[0]^2] \text{ partial}_{(2)}[g][0] \text{ partial}_{(1,1)}[g][0] +$
$2 f'[g[0]^2] \text{ partial}_{(1)}[g][0] \text{ partial}_{(1,2)}[g][0] + 4 g[0]^2 f''[g[0]^2]$
$\text{partial}_{(1)}[g][0] \text{ partial}_{(1,2)}[g][0] + g[0] f'[g[0]^2] \text{ partial}_{(1,1,2)}[g][0] \left. \right) +$
$27 x_1 x_2^2 \left( 6 g[0] f''[g[0]^2] \text{ partial}_{(1)}[g][0] \text{ partial}_{(2)}[g][0]^2 + \right.$
$4 g[0]^3 f^{(3)}[g[0]^2] \text{ partial}_{(1)}[g][0] \text{ partial}_{(2)}[g][0]^2 +$
$2 f'[g[0]^2] \text{ partial}_{(2)}[g][0] \text{ partial}_{(1,2)}[g][0] +$
$4 g[0]^2 f''[g[0]^2] \text{ partial}_{(2)}[g][0] \text{ partial}_{(1,2)}[g][0] +$
$f'[g[0]^2] \text{ partial}_{(1)}[g][0] \text{ partial}_{(2,2)}[g][0] + 2 g[0]^2 f''[g[0]^2]$
$\text{partial}_{(1)}[g][0] \text{ partial}_{(2,2)}[g][0] + g[0] f'[g[0]^2] \text{ partial}_{(1,2,2)}[g][0] \left. \right) +$
$9 x_2^3 \left( 6 g[0] f''[g[0]^2] \text{ partial}_{(2)}[g][0]^3 + 4 g[0]^3 f^{(3)}[g[0]^2] \text{ partial}_{(2)}[g][0]^3 + \right.$
$3 f'[g[0]^2] \text{ partial}_{(2)}[g][0] \text{ partial}_{(2,2)}[g][0] + 6 g[0]^2 f''[g[0]^2]$
$\text{partial}_{(2)}[g][0] \text{ partial}_{(2,2)}[g][0] + g[0] f'[g[0]^2] \text{ partial}_{(2,2,2)}[g][0] \left. \right)$

## Vector-valued functions

Suppose we compute the partial derivative with respect to $x_1$ of the function that takes $x$ to $f\big(g\big(h(x)^2\big)\big)$:

*In[44]:=* `partial[f[g[h[x]²]], x₁]`

*Out[44]=* $2 h[x] f'[g[h[x]^2]] g'[h[x]^2] \text{ partial}_{(1)}[h][x]$

The example above illustrates a general principle: This package assumes that all functions are real valued, unless told otherwise. Thus the last output is correct if we are thinking of $g$ as a real-valued function. If we want to think of $g$ as taking values in $\mathbf{R}^n$, we must first enter the command **setDimension[g[_], n]**, which instructs the computer that $g$ is a function with range in $\mathbf{R}^n$:

*In[45]:=*

> **setDimension[g[_], n]**
>
> ::: setDimension:
>
> g will be considered to be a function taking values in n−dimensional Euclidean space.

*In[46]:=* `partial[f[g[h[x]²]], x₁]`

*Out[46]=* $2 \, \text{gradient}[\text{f}]\big[\text{g}[\text{h}[x]^2]\big] \, . \, \big(\text{h}[x] \, \text{g}'[\text{h}[x]^2] \, \text{partial}_{(1)}[\text{h}][x]\big)$

Note that the last output involves the dot product of two vectors: the gradient of $f$, evaluated at $g(h(x)^2)$, and the derivative of $g$, evaluated at $h(x)^2$. The last two outputs should be compared.

## volume

**volume[n]** is the volume (unnormalized) of the unit ball in $\mathbf{R}^n$:

*In[47]:=* `volume[4]`

*Out[47]=* $\dfrac{\pi^2}{2}$

**volume** is computed using the formula given by Exercise 6 in Appendix A of [ABR].

## surfaceArea

**surfaceArea[n]** is the surface area (unnormalized) of the unit sphere in $\mathbf{R}^n$:

*In[48]:=* `surfaceArea[57]`

*Out[48]=* $\dfrac{536\,870\,912 \, \pi^{28}}{8\,687\,364\,368\,561\,751\,199\,826\,958\,100\,282\,265\,625}$

**surfaceArea[n]** is computed by multiplying the formula for **volume[n]** by $n$; see A.2 in

Appendix A of [ABR].

---

## integrateSphere

**integrateSphere[f, x]** equals the integral of $f$, with respect to normalized surface area measure, over the unit sphere in the Euclidean space defined by $x$. Here $f$ should be a polynomial function of $x$:

*In[49]:=* 
```
setDimension[x, n]
```

    ⋯   setDimension:

     x will be considered to be a vector in n−dimensional real Euclidean space.

*In[50]:=* 
```
integrateSphere[x₁² x₂⁴ x₃⁶, x]
```

*Out[50]=* 
$$\frac{45}{n \, (2+n) \, (4+n) \, (6+n) \, (8+n) \, (10+n)}$$

**integrateSphere[f, x]** differs from using *Mathematica*'s **Integrate** with *Mathematica*'s **Sphere** because **integrateSphere** allows the dimension of **x** to be either a symbol (as in the example above) or a concrete number, but *Mathematica*'s **Integrate** with *Mathematica*'s **Sphere** can compute integrations only when the dimension is a concrete number (such as 3). Also, integrateSphere uses normalized surface area measure, but *Mathematica*'s **Integrate** with *Mathematica*'s **Sphere** uses unnormalized surface area measure.

**integrateSphere** is computed by using the results in Section 3 of Hermann Weyl's paper [W].

---

## integrateBall

**integrateBall[f, x]** equals the integral of $f$, with respect to (unnormalized) volume measure, over the unit ball in the Euclidean space defined by $x$. Here $f$ should be a function of $x$ and $\|x\|$ that is a polynomial in $x$ and a function in $\|x\|$ for which *Mathematica* can find an explicit antiderivative:

*In[51]:=*  `setDimension[x, 7]`

    ⋯ setDimension:

       x will be considered to be a vector in 7−dimensional real Euclidean space.

*In[52]:=*  `integrateBall[x₁² x₂⁴ / (1 + ‖x‖), x]`

*Out[52]=*  $\dfrac{16\,\pi^3\,\left(-\dfrac{18107}{27720}+\text{Log}[2]\right)}{3465}$

`integrateBall[f, x]` differs from using *Mathematica*'s `Integrate` with *Mathematica*'s `Ball` because `integrateBall` allows the dimension of *x* to be either a symbol (as in the example below) or a concrete number (as in the example above), but *Mathematica*'s `Integrate` with *Mathematica*'s `Sphere` can compute integrations only when the dimension is a concrete number (such as 3). Another difference is that `integrateSphere` can deal with expressions such as ‖x‖ in the integrand (as in the examples above and below), and `integrateSphere` knows that $x_j$ denotes the $j$th-coordinate of *x* (as in the examples above and below).

The following expression finds the norm of $x_1\,x_2{}^4$ in the space $L^2\left(B,\ \left(1-\|x\|^2\right)dV\right)$, where *B* is the unit ball in $\mathbf{R}^n$ and dV is (unnormalized) volume measure on *B*.

*In[53]:=*  `setDimension[x, n]`

    ⋯ setDimension:

       x will be considered to be a vector in n−dimensional real Euclidean space.

*In[54]:=*  $\sqrt{\text{integrateBall}\left[\left(x_1\,x_2{}^4\right)^2\left(1-\|x\|^2\right),\,x\right]}$

*Out[54]=*  $\sqrt{210}\ \sqrt{\dfrac{\text{volume}[n]}{(2+n)\,(4+n)\,(6+n)\,(8+n)\,(10+n)\,(12+n)}}$

**integrateBall** is computed by converting to polar coordinates and then using the function **integrateSphere**  (see [R], Chapter 8, Exercise 6).

# integrateEllipsoidArea

**integrateEllipsoidArea[f, b, c, d, x]** equals the integral with respect to surface area measure of $\dfrac{f(x)}{\|(\nabla q)(x)\|}$ over the ellipsoid $\{x \in \mathbf{R}^n : q(x) = 0\}$, where $n$ equals the dimension of $x$ and $q(x) = b.x^2 + c.x + d$. In other words, **integrateEllipsoidArea[f, b, c, d, x]** equals the integral of

$$\frac{f(x)}{\sqrt{4\,b_1{}^2\,x_1{}^2 + 4\,b_1\,c_1\,x_1 + c_1{}^2 + \ldots + 4\,b_n{}^2\,x_n{}^2 + 4\,b_n\,c_n\,x_n + c_n{}^2}}$$

over the ellipsoid $\{x \in \mathbf{R}^n : q(x) = 0\}$. Here $f$ should be a polynomial function of $x$. Note that this integral is computed with respect to genuine surface area measure, not with surface area normalized to have total surface area 1 (as is the case with **integrateSphere**).

For example, we can compute the integral of $\dfrac{x_1{}^2\,x_2{}^6\,x_3{}^5}{\|(\nabla q)(x)\|}$ over the ellipsoid $\{x \in \mathbf{R}^3 : q(x) = 0\}$, where $q(x) = x_1{}^2 + 4\,x_2{}^2 + 3\,x_3{}^2 + 5\,x_1 + x_2 - 2\,x_3 - 6$, as follows:

*In[55]:=*    **setDimension[x, 3]**

⋯ setDimension:
    x will be considered to be a vector in 3-dimensional real Euclidean space.

*In[56]:=*    **integrateEllipsoidArea[x₁² x₂⁶ x₃⁵, {1, 4, 3}, {5, 1, -2}, -6, x]**

*Out[56]=*    $\dfrac{29\,949\,695\,715\,392\,943\,781\,937\,\sqrt{607}\,\pi}{54\,835\,301\,675\,238\,948\,864}$

Using the notation above, if $c$ consists of all 0's, then it may be omitted. Similarly, if $d = -1$, then $d$ may be omitted (independently of whether or not $c$ has been omitted). For example, we have the following:

*In[57]:=*    **integrateEllipsoidArea[x₁² x₂⁶ x₃⁴, {1, 4, 3}, x]**

*Out[57]=*    $\dfrac{\pi}{1\,729\,728\,\sqrt{3}}$

*In[58]:=*    **integrateEllipsoidArea[x₁⁸, {1, 4, 3}, x]**

*Out[58]=*    $\dfrac{\pi}{9\,\sqrt{3}}$

*In[59]:=* `integrateEllipsoidArea[9 x₁^8 - 1 729 728 x₁² x₂⁶ x₃⁴, {1, 4, 3}, x]`

*Out[59]=* `0`

Because the last result equals 0, the function $\left(9\,x_1{}^8 - 1\,729\,728\,x_1{}^2\,x_2{}^6\,x_3{}^4\right)\big/\|\nabla(x_1{}^2 + 4\,x_2{}^2 + 3\,x_3{}^2)\|$
[which equals $\left(9\,x_1{}^8 - 1\,729\,728\,x_1{}^2\,x_2{}^6\,x_3{}^4\right)\big/\left(2\,\sqrt{x_1{}^2 + 16\,x_2{}^2 + 9\,x_3{}^2}\,\right)$] is the normal derivative on
the ellipsoid $x_1{}^2 + 4\,x_2{}^2 + 3\,x_3{}^2 = 1$ of some harmonic polynomial on $\mathbf{R}^3$ (see Theorem 2.2 of [AS]).

**integrateEllipsoidArea[p, b, c, d, x]** is computed by using Proposition 3.2 in
[AS].

## integrateEllipsoidVolume

**integrateEllipsoidVolume[f, b, c, d, x]** equals the integral of $f$ with respect to
volume measure over $\{x \in \mathbf{R}^n : q(x) < 0\}$, where $n$ equals the dimension of $x$ and $q(x) =$
$b.x^2 + c.x + d$. Here $f$ should be a polynomial function of $x$. For example, we can compute the
integral of $x_1{}^2\,x_2{}^6\,x_3{}^5$ over $\left\{x \in \mathbf{R}^3 : x_1{}^2 + 4\,x_2{}^2 + 3\,x_3{}^2 + 5\,x_1 + x_2 - 2\,x_3 < 6\right\}$ as follows:

*In[60]:=* `integrateEllipsoidVolume[ x₁² x₂⁶ x₃⁵, {1, 4, 3}, {5, 1, -2}, -6, x]`

*Out[60]=* $\dfrac{894\,963\,845\,974\,894\,457\,\sqrt{607}\,\pi}{129\,818\,422\,526\,607\,360}$

Using the notation above, if $c$ consists of all 0's, then it may be omitted. Similarly, if $d = -1$, then $d$
may be omitted (independently of whether or not $c$ has been omitted):

*In[61]:=* `integrateEllipsoidVolume[ (w.x)⁴, {1, 4, 3}, x]`

*Out[61]=* $\dfrac{\pi\,\left(144\,w_1^4 + 72\,w_1^2\,w_2^2 + 9\,w_2^4 + 96\,w_1^2\,w_3^2 + 24\,w_2^2\,w_3^2 + 16\,w_3^4\right)}{2520\,\sqrt{3}}$

# Boundary Value Problems

## Dirichlet problems

### *The Dirichlet problem on the sphere*

**dirichlet[p, x]** is the solution to the standard Dirichlet problem: find the harmonic function on the unit ball in the Euclidean space defined by $x$ that equals $p$ on the unit sphere. Thus **dirichlet[p, x]** is the Poisson integral of $p$ as a function of $x$. Here $p$ must be a polynomial function of $x$:

*In[62]:=*
```
setDimension[x, 5]
```

··· setDimension:
   x will be considered to be a vector in 5-dimensional real Euclidean space.

*In[63]:=*
```
dirichlet[x₁⁴ x₂², x]
```

*Out[63]=*
$$\frac{1}{15\,015}\left(143 - 273\,\|x\|^2 + 165\,\|x\|^4 - 35\,\|x\|^6 + 910\,x_1^2 - 1540\,\|x\|^2\,x_1^2 + 630\,\|x\|^4\,x_1^2 + 1155\,x_1^4 - 1155\,\|x\|^2\,x_1^4 + 455\,x_2^2 - 770\,\|x\|^2\,x_2^2 + 315\,\|x\|^4\,x_2^2 + 6930\,x_1^2\,x_2^2 - 6930\,\|x\|^2\,x_1^2\,x_2^2 + 15\,015\,x_1^4\,x_2^2\right)$$

The last output is a harmonic function on $\mathbf{R}^5$ that equals $x_1^4\,x_2^2$ on the unit sphere. To check this, first we take the Laplacian of the last output, and then we evaluate the last output on the unit sphere:

*In[64]:=*
```
Δₓ[%]
```

*Out[64]=*
```
0
```

*In[65]:=*
```
%% /. ‖x‖ → 1
```

*Out[65]=*
$$x_1^4\,x_2^2$$

Thus we have indeed solved the specified Dirichlet problem.

The solution to the Dirichlet problem on the sphere is computed by using the algorithm described in [AR].

### The Dirichlet problem on the exterior of the sphere

The `region` option allows the user to solve Dirichlet problems on regions other than the sphere. Currently supported values for `region` are `Sphere`, `exteriorSphere`, `annulus`, and `quadratic`. The default value of `region` is `Sphere`; thus `dirichlet[p, x, region → Sphere]` is the same as `dirichlet[p, x]`.

`dirichlet[p, x, region → exteriorSphere]` gives the solution to the standard exterior Dirichlet problem: find the harmonic function on the exterior (including $\infty$) of the unit ball in the Euclidean space defined by $x$ that equals $p$ on the unit sphere. This function is the exterior Poisson integral of $p$ as a function of $x$ (see Theorem 4.11 of [ABR]). Here $p$ must be a polynomial function of $x$:

*In[66]:=* `dirichlet[x₁⁴ x₂², x, region → exteriorSphere]`

*Out[66]=* $\frac{1}{15\,015\,\|x\|^{15}} \left(-35\,\|x\|^6 + 165\,\|x\|^8 - 273\,\|x\|^{10} + 143\,\|x\|^{12} + 630\,\|x\|^4\,x_1^2 - 1540\,\|x\|^6\,x_1^2 + 910\,\|x\|^8\,x_1^2 - 1155\,\|x\|^2\,x_1^4 + 1155\,\|x\|^4\,x_1^4 + 315\,\|x\|^4\,x_2^2 - 770\,\|x\|^6\,x_2^2 + 455\,\|x\|^8\,x_2^2 - 6930\,\|x\|^2\,x_1^2\,x_2^2 + 6930\,\|x\|^4\,x_1^2\,x_2^2 + 15\,015\,x_1^4\,x_2^2\right)$

The last output is a harmonic function on the exterior of the unit ball in $\mathbf{R}^5$ (because the dimension of $x$ was set to 5 earlier) that equals $x_1{}^4\,x_2{}^2$ on the unit sphere.

The solution to the Dirichlet problem on the exterior of the sphere is computed by taking the Kelvin transform of the solution that would have been obtained without the `region → exteriorSphere` option; see Chapter 4 of [ABR].

### The Dirichlet problem on annular regions

`dirichlet[{p, q}, x, region → annulus[r, s] ]` gives the solution to the annular Dirichlet problem: find the harmonic function on the annular region with inner radius $r$ and outer radius $s$ that equals $p$ on the sphere of radius $r$ and equals $q$ on the sphere of radius $s$. Here $p$ and $q$ must be polynomial functions of $x$:

*In[67]:=*  `dirichlet[{x₁³, x₃²}, x, region → annulus[1, 4]]`

*Out[67]=*  $-\frac{1024}{315}\left(-1+\frac{1}{\|x\|^3}\right)+\frac{1024\left(-\frac{1}{1024}+\frac{1}{\|x\|^5}\right)x_1}{2387}+$

$\frac{\left(-262\,144+\|x\|^9\right)\left(3\,\|x\|^2\,x_1-7\,x_1^3\right)}{1\,835\,001\,\|x\|^9}-\frac{16\,384\left(-1+\frac{1}{\|x\|^7}\right)\left(-\frac{\|x\|^2}{5}+x_3^2\right)}{16\,383}$

Now we check that the last output is harmonic on $\mathbf{R}^5$ (because the dimension of $x$ was set to 5 earlier), that it equals $x_1^3$ on the sphere of radius 1, and that it equals $x_3^2$ on the sphere of radius 4:

*In[68]:=*  `Δ_x [%]`

*Out[68]=*  `0`

*In[69]:=*  `%% /. ‖x‖ → 1`

*Out[69]=*  $x_1^3$

*In[70]:=*  `%%% /. ‖x‖ → 4`

*Out[70]=*  $x_3^2$

If $p = q$, then the shortcut `dirichlet[p, x, region → annulus[r, s] ]` can be used instead of `dirichlet[{p, q}, x, region → annulus[r, s] ]`.

The solution to the Dirichlet problem on an annulus is computed by using the techniques described in Chapter 10 of [ABR].

### The Dirichlet problem on quadratic surfaces

`dirichlet[p, x, region → quadratic[b, c, d] ]` gives the harmonic polynomial that equals $p$ on the quadratic surface

$$b.x^2 + c.x + d = 0.$$

Here $b$ and $c$ must be lists whose length equals the dimension of $x$, and $p$ must be a polynomial function of $x$. For example, here we find the harmonic polynomial that equals $x_1{}^3 x_3{}^2$ on the elliptic paraboloid given by the equation

$$5 x_1{}^2 + 3 x_2{}^2 = 4 x_3 + 1$$

in $\mathbf{R}^3$:

*In[71]:=*    `setDimension[x, 3]`

     **···** <span style="color:red">setDimension</span>:

     x will be considered to be a vector in 3–dimensional real Euclidean space.

*In[72]:=*    `dirichlet[x₁³ x₃², x, region → quadratic[{5, 3, 0}, {0, 0, -4}, -1]]`

*Out[72]=*    $x_1^3 x_3^2 + \dfrac{1}{34\,506} \left(-1 + 5 x_1^2 + 3 x_2^2 - 4 x_3\right) \left(-1163 x_1 - 162 x_1^3 + 567 x_1 x_2^2 - 3816 x_1 x_3 - 5751 x_1 x_3^2\right)$

The last output is of the form $x_1{}^3 x_3{}^2 + (5 x_1{}^2 + 3 x_2{}^2 - 4 x_3 - 1) f$, where $f$ is a polynomial. Thus the last output obviously equals $x_1{}^3 x_3{}^2$ on the elliptic paraboloid given by the equation $5 x_1{}^2 + 3 x_2{}^2 = 4 x_3 + 1$. Thus to verify that the last output is correct, we only need check that it is harmonic:

*In[73]:=*    `Δₓ[%]`

*Out[73]=*    `0`

The solution to the Dirichlet problem on a quadratic surface is computed by using the algorithm developed in [AGV]. That paper also shows that if each coordinate of $b$ is nonnegative and at least one coordinate of $b$ is nonzero, and if the quadratic surface in question is nontrivial in the sense that there exits $x \in \mathbf{R}^n$ such that

$$b.x^2 + c.x + d < 0,$$

then the polynomial produced by this algorithm is the unique harmonic polynomial that equals $p$ on the quadratic surface in question.

Using the notation above, if $c$ consists of all 0's, then it may be omitted. For example, here we find the harmonic polynomial that equals $x_1{}^4 x_3{}^2$ on the ellipsoid given by the equation $7 x_1{}^2 + 3 x_2{}^2 + 4 x_3{}^2 = 1$:

*In[74]:=*     `dirichlet[x₁⁴ x₃², x, region → quadratic[{7, 3, 4}, -1]]`

*Out[74]=*     $x_1^4 \, x_3^2 + \left( \left( -1 + 7 \, x_1^2 + 3 \, x_2^2 + 4 \, x_3^2 \right) \right.$
         $\left( -2\,366\,781 - 22\,817\,375 \, x_1^2 - 41\,112\,960 \, x_1^4 + 6\,343\,407 \, x_2^2 + 57\,632\,526 \, x_1^2 \, x_2^2 - 5\,172\,930 \, x_2^4 - \right.$
         $\left. \left. 16\,660\,966 \, x_3^2 - 1\,001\,229\,054 \, x_1^2 \, x_3^2 + 38\,928\,834 \, x_2^2 \, x_3^2 + 54\,988\,584 \, x_3^4 \right) \right) / 11\,209\,827\,216$

The last output is of the form $x_1^4 \, x_3^2 + (7 \, x_1^2 + 3 \, x_2^2 + 4 \, x_3^2 - 1) \, f$, where $f$ is a polynomial. Thus the last output obviously equals $x_1^4 \, x_3^2$ on the ellipsoid given by the equation $7 \, x_1^2 + 3 \, x_2^2 + 4 \, x_3^2 = 1$. Thus to verify that the last output is correct, we only need check that it is harmonic:

*In[75]:=*     `Δₓ[%]`

*Out[75]=*     `0`

Using the notation above, if $d = -1$, then $d$ may be omitted. Thus
`dirichlet[x₁³ x₃², x, region → quadratic[{5, 3, 0}, {0, 0, -4}]]` is the same as
`dirichlet[x₁³ x₃², x, region → quadratic[{5, 3, 0}, {0, 0, -4}, -1]]`. Furthermore,
`dirichlet[x₁⁴ x₃², x, region → quadratic[{7, 3, 4}]]` is the same as
`dirichlet[x₁⁴ x₃², x, region → quadratic[{7, 3, 4}, -1]]` and
`dirichlet[x₁⁴ x₃², x, region → quadratic[{7, 3, 4}, {0, 0, 0}, -1]]`.

### Generalized Dirichlet problems

The generalized Dirichlet problem asks for a function with specified boundary values on some surface and with specified Laplacian inside that surface. Thus if the Laplacian is required to be 0, then this is the same as the usual Dirichlet problem. The generalized Dirichlet problem with specified Laplacian equal to $q$ is solved by adding the option $\Delta \to q$ to the `dirichlet` function; here $q$ must be a polynomial. For example, we can find the function that equals $x_1^3 \, x_2^2$ on the unit sphere in $\mathbf{R}^3$ and whose Laplacian equals $x_2^2 \, x_3$:

*In[76]:=*     `dirichlet[x₁³ x₂², x, Δ → x₂² x₃]`

*Out[76]=*     $\frac{1}{1260} \left( 108 \, x_1 - 168 \, \|x\|^2 \, x_1 + 60 \, \|x\|^4 \, x_1 + 140 \, x_1^3 - 140 \, \|x\|^2 \, x_1^3 + 420 \, x_1 \, x_2^2 - \right.$
         $\left. 420 \, \|x\|^2 \, x_1 \, x_2^2 + 1260 \, x_1^3 \, x_2^2 - 9 \, x_3 + 14 \, \|x\|^2 \, x_3 - 5 \, \|x\|^4 \, x_3 - 70 \, x_2^2 \, x_3 + 70 \, \|x\|^2 \, x_2^2 \, x_3 \right)$

We can check that output above correctly solves the generalized Dirichlet problem by verifying that the Laplacian of the last output indeed equals $x_2{}^2 x_3$ and that the last output restricted to the unit sphere indeed equals $x_1{}^3 x_2{}^2$:

*In[77]:=*    `Δ_x [%]`

*Out[77]=*    $x_2^2 x_3$

*In[78]:=*    `%% /. ‖x‖ → 1`

*Out[78]=*    $x_1^3 x_2^2$

The `Δ` option can be used combined with the **`region`** option to solve the generalized Dirichlet problem on any region for which this package can solve the usual Dirichlet problem. The order of the options does not matter. Here we find the function on $\mathbf{R}^3$ whose Laplacian equals $x_2{}^2$ and that equals $x_1{}^4 x_3{}^2$ on the ellipsoid $2 x_1{}^2 + 3 x_2{}^2 + 4 x_3{}^2 - 1 = 0$:

*In[79]:=*    `dirichlet[x_1^4 x_3^2, x, Δ → x_2^2, region → quadratic[{2, 3, 4}, -1]]`

*Out[79]=*    $x_1^4 x_3^2 +$
$\left( \left( -1 + 2 x_1^2 + 3 x_2^2 + 4 x_3^2 \right) \left( 82\,719\,889 - 2\,354\,066\,130\, x_1^2 - 3\,207\,153\,660\, x_1^4 + 3\,779\,085\,195\, x_2^2 + \right.\right.$
$3\,780\,731\,736\, x_1^2 x_2^2 - 270\,421\,740\, x_2^4 - 680\,540\,064\, x_3^2 - 23\,067\,975\,984\, x_1^2 x_3^2 +$
$\left.\left. 816\,437\,844\, x_2^2 x_3^2 + 1\,369\,325\,424\, x_3^4 \right) \right) / 157\,239\,173\,916$

The form of the last output shows, without calculation, that it obviously equals $x_1{}^4 x_3{}^2$ on the ellipsoid $2 x_1{}^2 + 3 x_2{}^2 + 4 x_3{}^2 - 1 = 0$. Thus to verify that we have correctly solved this generalized Dirichlet problem, we only need to check that the Laplacian equals $x_2{}^2$:

*In[80]:=*    `Δ_x [%]`

*Out[80]=*    $x_2^2$

The generalized Dirichlet problem on some region requires finding a function *u* that equals *p* on the boundary of the region and such that $\Delta u = q$. This problem is solved as follows: First find any function *v* such that $\Delta v = q$ (see the **`antiLaplacian`** section later in this document for the computation of this step). Then solve the usual Dirichlet problem for the specified region with boundary function $p - v$, getting a harmonic function $w$ such that $w$ equals $p - v$ on the boundary of the region. Now set $u = w + v$, getting the desired function $u$ that equals $p$ on the boundary of the region and such that $\Delta u = q$.

### Dirichlet problems with explicit coordinates

So far we have worked with symbolic vectors, which we have usually called $x$ or $z$. Vectors can also be described by giving explicit coordinates in the form of a list. Both formats (symbolic vectors and explicit lists) work with all functions in this package. For example, suppose we are working in $\mathbf{R}^3$ and want to use $x$, $y$, and $z$ instead of $x_1$, $x_2$, and $x_3$. Here is how to find the Poisson integral of $x^3 y z^2$:

*In[81]:=*
```
dirichlet[x^3 y z^2, {x, y, z}]
```

*Out[81]=*
$$\frac{1}{231} \left( 11 x y + 3 x^3 y - 14 x^5 y - 18 x y^3 - 7 x^3 y^3 + 7 x y^5 + 45 x y z^2 + 161 x^3 y z^2 - 49 x y^3 z^2 - 56 x y z^4 \right)$$

To check that the last output is correct, first we verify that it is harmonic by checking that its Laplacian equals 0:

*In[82]:=*
```
Δ_{x,y,z} [%]
```

*Out[82]=*
```
0
```

Now we check that the difference between the claimed solution to the Dirichlet problem equals $x^3 yz^2$ on the unit sphere by verifying that the difference between these two functions, when restricted to the unit sphere, equals 0:

*In[83]:=*
```
x^3 y z^2 - %% /. z → √(1 - x^2 - y^2)
```

*Out[83]=*
```
0
```

Here is another example using explicit coordinates. In this example, we find the function on $\mathbf{R}^3$ whose Laplacian equals $y^2 z$ and that equals $x^3 y z^2$ on the ellipsoid $2 x^2 + 3 y^2 + 4 z^2 - 1 = 0$:

*In[84]:=*
```
dirichlet[x^3 y z^2, {x, y, z}, region → quadratic[{2, 3, 4}, -1], Δ → y^2 z^3]
```

*Out[84]=*

$x^3 y z^2 + (-1 + 2 x^2 + 3 y^2 + 4 z^2)$

$\left( -\dfrac{961 x y}{150\,917} - \dfrac{223 x^3 y}{7943} + \dfrac{93 x y^3}{7943} + \dfrac{6\,128\,862\,523 z}{154\,091\,469\,268\,302} - \dfrac{424\,717\,558 x^2 z}{4\,532\,102\,037\,303} + \right.$

$\dfrac{44\,102 x^4 z}{734\,657\,487} + \dfrac{3\,416\,529\,117 y^2 z}{6\,042\,802\,716\,404} - \dfrac{448\,445 x^2 y^2 z}{489\,771\,658} - \dfrac{1\,320\,573 y^4 z}{979\,543\,316} -$

$\dfrac{571 x y z^2}{7943} + \dfrac{1\,235\,669\,309 z^3}{18\,128\,408\,149\,212} - \dfrac{213\,085 x^2 z^3}{1\,469\,314\,974} + \dfrac{8\,956\,053 y^2 z^3}{979\,543\,316} - \left. \dfrac{297\,101 z^5}{734\,657\,487} \right)$

Obviously the last output equals $x^3 y z^2$ on the ellipsoid $2 x^2 + 3 y^2 + 4 z^2 - 1 = 0$. We now check that its Laplacian is the desired function:

*In[85]:=*
```
Δ_{x,y,z}[%]
```

*Out[85]=*

$y^2 z^3$

---

## harmonicDecomposition

**harmonicDecomposition[p, x]** gives the decomposition of $p$ into a sum of harmonic polynomials (on the Euclidean space defined by $x$) times even powers of $\|x\|$. Here $p$ must be a polynomial function of $x$:

*In[86]:=*
```
setDimension[x, n]
```

··· setDimension:

x will be considered to be a vector in n−dimensional real Euclidean space.

*In[87]:=*
```
harmonicDecomposition[x_1^4, x]
```

*Out[87]=*

$\left\{ \left\{ \dfrac{3 \|x\|^4}{(2 + n)(4 + n)} - \dfrac{6 \|x\|^2 x_1^2}{4 + n} + x_1^4, 0 \right\}, \left\{ \dfrac{6 (-\|x\|^2 + n x_1^2)}{n(4 + n)}, 2 \right\}, \left\{ \dfrac{3}{n(2 + n)}, 4 \right\} \right\}$

The output given by **harmonicDecomposition[p, x]** consists of a list of pairs. The first entry in each pair is a harmonic function, the second entry is the power of $\|x\|$ by which the first entry should be multiplied so that the sum of the resulting terms equals $p$. For example, in the last output

$$\frac{3\|x\|^4}{(2+n)(4+n)} - \frac{6\|x\|^2 x_1^2}{4+n} + x_1^4 \text{ and } \frac{6\left(-\|x\|^2 + n\,x_1^2\right)}{n(4+n)} \text{ and } \frac{3}{n(2+n)} \text{ are harmonic functions of } x;$$

furthermore

$$x_1^4 = \frac{3\|x\|^4}{(2+n)(4+n)} - \frac{6\|x\|^2 x_1^2}{4+n} + x_1^4 + \frac{6\left(-\|x\|^2 + n\,x_1^2\right)}{n(4+n)}\|x\|^2 + \frac{3}{n(2+n)}\|x\|^4.$$

For a proof of the existence and uniqueness of the harmonic decomposition, see [ABR], Theorem 5.7. **harmonicDecomposition** is computed using the algorithm described in Section 2 of [AR].

---

## antiLaplacian

**antiLaplacian[u, x]** gives an anti-Laplacian of *u* with respect to *x*; this is a function whose Laplacian with respect to *x* equals *u*. Here *u* must be a polynomial function of *x* or a sum of terms, each of which is a polynomial in *x* times a function of $\|x\|$:

*In[88]:=*  **setDimension[x, 5]**

⋯ setDimension:
    x will be considered to be a vector in 5−dimensional real Euclidean space.

*In[89]:=*  **antiLaplacian[$x_1^2\,x_2\,\|x\|^3\,$Log[$\|x\|$], x]**

*Out[89]=*  $\dfrac{\left(-19\,\|x\|^7 + 84\,\text{Log}[\|x\|]\,\|x\|^7\right)x_2}{49\,392} - \dfrac{\left(-19\,\|x\|^5 + 70\,\text{Log}[\|x\|]\,\|x\|^5\right)\left(\|x\|^2 x_2 - 7\,x_1^2\,x_2\right)}{34\,300}$

The function given by the last output has Laplacian equal to $x_1^2\,x_2\,\|x\|^3\,\log(\|x\|)$ on $\mathbf{R}^5$, as we can easily verify:

*In[90]:=*  **Δ$_x$[%]**

*Out[90]=*  Log[$\|x\|$] $\|x\|^3\,x_1^2\,x_2$

Note that finding the anti-Laplacian of even a fairly simple function such as $x_1^2\,x_2\,\|x\|^3\,\log(\|x\|)$ led to a solution containing five-digit integers.

Here is another interesting example and its verification. Note in the example below that the anti-Laplacian is unbounded near 0, even though $x_1 x_2 \cos(\|x\|)$ is bounded near 0. However, this behavior is not surprising because $\cos(\|x\|)$ is not smooth at 0:

*In[91]:=* 
```
antiLaplacian[ x₁ x₂ Cos[∥x∥] , x]
```

*Out[91]=* 
$$\frac{1}{\|x\|^7} \left( -\text{Cos}[\|x\|] \ \|x\| \ \left( -5760 + 960 \ \|x\|^2 - 48 \ \|x\|^4 + \|x\|^6 \right) + 8 \left( -720 + 360 \ \|x\|^2 - 30 \ \|x\|^4 + \|x\|^6 \right) \text{Sin}[\|x\|] \right) x_1 x_2$$

*In[92]:=* 
```
Δₓ[%]
```

*Out[92]=* 
```
Cos[∥x∥] x₁ x₂
```

To understand how **antiLaplacian** is computed, suppose we want to find an anti-Laplacian of $p(x)f(\|x\|)$, where $p$ is a polynomial on $\mathbf{R}^n$ and $f$ is a continuous function on $(0, \infty)$. Applying **harmonicDecomposition** to $p$, using linearity, and replacing $f(\|x\|)$ with an even power of $\|x\|$ times $f(\|x\|)$, we see that we only need to find anti-Laplacians of functions of the form $q(x)f(\|x\|)$, where $q$ is a harmonic polynomial on $\mathbf{R}^n$ homogeneous of degree $m$. To find an anti-Laplacian of $q(x)f(\|x\|)$, suppose $h$ is a twice-differentiable function on $(0, \infty)$. The Laplacian of $q(x)h(\|x\|)$ equals

$$q(x) \ \frac{(2\,m + n - 1) \ h'(\|x\|) + \|x\| \ h''(\|x\|)}{\|x\|},$$

where we have used the product rule for Laplacians (see 1.19 in [ABR]) and Exercise 29, Chapter 1 of [ABR]. We want to make the expression above equal $q(x)f(\|x\|)$, so we need only solve the differential equation

$$\frac{(2\,m + n - 1) \ h'(t) + t \ h''(t)}{t} \ = \ f(t).$$

A solution to this equation is given by

$$h(t) \ = \ \int_a^t s^{1-2\,m-n} \int_b^s r^{2\,m+n-1} \ f(r) \ dr \ ds$$

for any choice of the constants $a$ and $b$. **antiLaplacian** uses the formula above with $b = 0$, although if a singularity is encountered at 0 it automatically makes a different choice. If you know there is a singularity at 0, you can save some time by using the option **singularity → 0**, as illustrated by the following example. The two timing tests below show a large speed-up when using the option **singularity → 0**:

*In[93]:=* 
```
Timing[antiLaplacian[x₁² x₂ Log[‖x‖]¹⁰, x]]
```

*Out[93]=* 
$\Big\{4.125, \frac{1}{44\,432\,481\,042\,432}$

$\big(1\,098\,190\,285\,675\, \|x\|^4 - 4\,392\,027\,139\,500\, \text{Log}[\|x\|]\, \|x\|^4 + 8\,780\,751\,264\,600\, \text{Log}[\|x\|]^2\, \|x\|^4 - $

$11\,697\,759\,309\,600\, \text{Log}[\|x\|]^3\, \|x\|^4 + 11\,675\,463\,962\,400\, \text{Log}[\|x\|]^4\, \|x\|^4 - $

$9\,300\,239\,544\,960\, \text{Log}[\|x\|]^5\, \|x\|^4 + 6\,139\,962\,259\,200\, \text{Log}[\|x\|]^6\, \|x\|^4 - $

$3\,431\,153\,157\,120\, \text{Log}[\|x\|]^7\, \|x\|^4 + 1\,628\,505\,285\,120\, \text{Log}[\|x\|]^8\, \|x\|^4 - $

$636\,708\,833\,280\, \text{Log}[\|x\|]^9\, \|x\|^4 + 176\,319\,369\,216\, \text{Log}[\|x\|]^{10}\, \|x\|^4\big)\, x_2 - $

$\frac{1}{15\,977\,453\,554\,216}\big(449\,365\,877\,986\,725\, \|x\|^2 - 898\,731\,726\,943\,050\, \text{Log}[\|x\|]\, \|x\|^2 + $

$898\,731\,567\,275\,850\, \text{Log}[\|x\|]^2\, \|x\|^2 - 599\,153\,792\,737\,500\, \text{Log}[\|x\|]^3\, \|x\|^2 + $

$299\,575\,286\,391\,150\, \text{Log}[\|x\|]^4\, \|x\|^2 - 119\,826\,572\,605\,740\, \text{Log}[\|x\|]^5\, \|x\|^2 + $

$39\,935\,697\,292\,260\, \text{Log}[\|x\|]^6\, \|x\|^2 - 11\,399\,995\,035\,000\, \text{Log}[\|x\|]^7\, \|x\|^2 + $

$2\,835\,967\,995\,630\, \text{Log}[\|x\|]^8\, \|x\|^2 - 613\,066\,399\,660\, \text{Log}[\|x\|]^9\, \|x\|^2 + $

$103\,749\,698\,404\, \text{Log}[\|x\|]^{10}\, \|x\|^2\big)\,\big(\|x\|^2 x_2 - 7\, x_1^2\, x_2\big)\Big\}$

*In[94]:=* 
```
Timing[antiLaplacian[x₁² x₂ Log[‖x‖]¹⁰, x, singularity → 0]]
```

*Out[94]=* 
$\Big\{0.046875, \frac{1}{44\,432\,481\,042\,432}$

$\big(1\,098\,190\,285\,675\, \|x\|^4 - 4\,392\,027\,139\,500\, \text{Log}[\|x\|]\, \|x\|^4 + 8\,780\,751\,264\,600\, \text{Log}[\|x\|]^2\, \|x\|^4 - $

$11\,697\,759\,309\,600\, \text{Log}[\|x\|]^3\, \|x\|^4 + 11\,675\,463\,962\,400\, \text{Log}[\|x\|]^4\, \|x\|^4 - $

$9\,300\,239\,544\,960\, \text{Log}[\|x\|]^5\, \|x\|^4 + 6\,139\,962\,259\,200\, \text{Log}[\|x\|]^6\, \|x\|^4 - $

$3\,431\,153\,157\,120\, \text{Log}[\|x\|]^7\, \|x\|^4 + 1\,628\,505\,285\,120\, \text{Log}[\|x\|]^8\, \|x\|^4 - $

$636\,708\,833\,280\, \text{Log}[\|x\|]^9\, \|x\|^4 + 176\,319\,369\,216\, \text{Log}[\|x\|]^{10}\, \|x\|^4\big)\, x_2 - $

$\frac{1}{15\,977\,453\,554\,216}\big(449\,365\,877\,986\,725\, \|x\|^2 - 898\,731\,726\,943\,050\, \text{Log}[\|x\|]\, \|x\|^2 + $

$898\,731\,567\,275\,850\, \text{Log}[\|x\|]^2\, \|x\|^2 - 599\,153\,792\,737\,500\, \text{Log}[\|x\|]^3\, \|x\|^2 + $

$299\,575\,286\,391\,150\, \text{Log}[\|x\|]^4\, \|x\|^2 - 119\,826\,572\,605\,740\, \text{Log}[\|x\|]^5\, \|x\|^2 + $

$39\,935\,697\,292\,260\, \text{Log}[\|x\|]^6\, \|x\|^2 - 11\,399\,995\,035\,000\, \text{Log}[\|x\|]^7\, \|x\|^2 + $

$2\,835\,967\,995\,630\, \text{Log}[\|x\|]^8\, \|x\|^2 - 613\,066\,399\,660\, \text{Log}[\|x\|]^9\, \|x\|^2 + $

$103\,749\,698\,404\, \text{Log}[\|x\|]^{10}\, \|x\|^2\big)\,\big(\|x\|^2 x_2 - 7\, x_1^2\, x_2\big)\Big\}$

In the case where **p** is a polynomial (so that there are no terms such as **Log[‖x‖]** or other functions of **‖x‖** as in the examples above), then **antiLaplacian[p, x]** is not computed using the procedure described above. Instead, the anti-Laplacian of each monomial is computed using a much faster iterative procedure that at each step increases the degree of the first coordinate and decreases the degree of other coordinates; see Lemma 2.7 in [AS] for this procedure, which always produces a polynomial for the anti-Laplacian. Here is an example:

*In[95]:=* `antiLaplacian[x₁² x₂⁵ + 6 x₁³ x₂² x₃⁴, x]`

*Out[95]=* $\frac{1}{42} x_1^2 x_2^7 - \frac{x_2^9}{1512} + \frac{420 x_1^3 x_2^2 x_3^6 - 15 x_1^3 x_3^8 - 45 x_1 x_2^2 x_3^8 + 2 x_1 x_3^{10}}{2100}$

*In[96]:=* `Δ_x[%]`

*Out[96]=* $x_1^2 x_2^5 + 6 x_1^3 x_2^2 x_3^4$

The anti-Laplacian of a given function is never unique. However, each polynomial has a unique anti-Laplacian that is a polynomial multiple of $\|x\|^2$, where $x$ is the variable. The option **multiple → norm²** produces the unique anti-Laplacian that is a polynomial multiple of $\|x\|^2$:

*In[97]:=* `antiLaplacian[x₁² x₂⁵, x, multiple → norm²]`

*Out[97]=* $\frac{1}{302328}$
$\left(-9 \|x\|^8 x_2 + 156 \|x\|^6 x_1^2 x_2 + 104 \|x\|^6 x_2^3 - 2340 \|x\|^4 x_1^2 x_2^3 - 234 \|x\|^4 x_2^5 + 7956 \|x\|^2 x_1^2 x_2^5\right)$

*In[98]:=* `Δ_x[%]`

*Out[98]=* $x_1^2 x_2^5$

The option **multiple → quadratic[b, c, d]** produces the unique anti-Laplacian that is a polynomial multiple of the quadratic expression $b.x^2 + c.x + d$. Here $b$ and $c$ are lists of numbers or symbols with length dimension[$x$], and $d$ is a number or symbol:

*In[99]:=* `setDimension[x, 3]`

⋯ setDimension:

x will be considered to be a vector in 3–dimensional real Euclidean space.

*In[100]:=*   `antiLaplacian[x₁² x₂ x₃, x, multiple → quadratic[{7, 3, 5}, {6, 4, 2}, -8]]`

*Out[100]=*

$$\frac{1}{1\,507\,708\,465\,430\,520\,600\,292\,500}$$
$$\left(-8 + 6\,x_1 + 7\,x_1^2 + 4\,x_2 + 3\,x_2^2 + 2\,x_3 + 5\,x_3^2\right)\ \left(447\,373\,820\,559\,267\,521\,408 - \right.$$
$$169\,551\,027\,034\,920\,871\,200\,x_1 + 52\,724\,035\,196\,061\,138\,000\,x_1^2 - 644\,786\,494\,897\,819\,357\,200\,x_2 + $$
$$458\,341\,693\,344\,640\,942\,200\,x_1\,x_2 - 422\,543\,240\,577\,614\,520\,000\,x_1^2\,x_2 + $$
$$60\,139\,948\,483\,932\,408\,000\,x_2^2 + 72\,016\,343\,331\,465\,487\,500\,x_2^3 - $$
$$998\,160\,853\,689\,847\,330\,560\,x_3 + 806\,274\,746\,937\,510\,606\,000\,x_1\,x_3 - $$
$$777\,532\,413\,963\,810\,960\,000\,x_1^2\,x_3 + 3\,172\,006\,238\,006\,936\,421\,000\,x_2\,x_3 - $$
$$4\,208\,313\,650\,329\,240\,247\,250\,x_1\,x_2\,x_3 + 11\,842\,617\,023\,843\,893\,048\,125\,x_1^2\,x_2\,x_3 - $$
$$601\,976\,778\,299\,284\,342\,500\,x_2^2\,x_3 - 549\,566\,395\,101\,035\,983\,125\,x_2^3\,x_3 + $$
$$26\,865\,198\,721\,070\,892\,000\,x_3^2 - 219\,307\,969\,328\,878\,316\,250\,x_2\,x_3^2 + $$
$$\left.166\,443\,532\,883\,241\,705\,000\,x_3^3 - 772\,266\,502\,919\,756\,446\,875\,x_2\,x_3^3\right)$$

*In[101]:=*   `Δₓ[%]`

*Out[101]=*   $x_1^2\,x_2\,x_3$

Using the notation above, if $c$ consists of all 0's, then it may be omitted. Similarly, if $d = -1$, then $d$ may be omitted (independently of whether or not $c$ has been omitted). For example, we have the following:

*In[102]:=*   `antiLaplacian[x₁² x₂⁵, x, multiple → quadratic[{5, 3, 2}]]`

*Out[102]=*

$$\frac{1}{581\,833\,767\,288\,446\,820\,864}$$
$$\left(-1 + 5\,x_1^2 + 3\,x_2^2 + 2\,x_3^2\right)\ \left(2\,456\,037\,114\,711\,717\,x_2 + 11\,849\,131\,274\,921\,369\,x_1^2\,x_2 - \right.$$
$$99\,364\,687\,683\,541\,365\,x_1^4\,x_2 + 161\,129\,212\,822\,880\,475\,x_1^6\,x_2 + 11\,647\,115\,153\,301\,463\,x_2^3 + $$
$$420\,073\,918\,355\,826\,754\,x_1^2\,x_2^3 - 1\,675\,238\,953\,996\,349\,345\,x_1^4\,x_2^3 + 6\,105\,788\,388\,568\,659\,x_2^5 + $$
$$3\,504\,622\,438\,227\,426\,081\,x_1^2\,x_2^5 - 90\,937\,993\,438\,762\,875\,x_2^7 - 7\,493\,882\,899\,438\,286\,x_2\,x_3^2 - $$
$$40\,981\,933\,892\,125\,428\,x_1^2\,x_2\,x_3^2 + 159\,614\,634\,738\,057\,690\,x_1^4\,x_2\,x_3^2 - 37\,122\,796\,442\,909\,964\,x_2^3\,x_3^2 - $$
$$742\,042\,463\,001\,143\,716\,x_1^2\,x_2^3\,x_3^2 - 18\,666\,023\,074\,849\,206\,x_2^5\,x_3^2 + 8\,515\,053\,550\,851\,900\,x_2\,x_3^4 + $$
$$\left.34\,172\,978\,315\,176\,644\,x_1^2\,x_2\,x_3^4 + 29\,420\,004\,609\,142\,012\,x_2^3\,x_3^4 - 3\,498\,085\,489\,788\,648\,x_2\,x_3^6\right)$$

*In[103]:=*   `Δₓ[%]`

*Out[103]=*   $x_1^2\,x_2^5$

`antiLaplacian` with the option `multiple → quadratic` is computed by using the algorithm developed in [AGV].

## neumann

`neumann[f, x]` is the solution to the standard Neumann problem: find the harmonic function on the unit ball in the Euclidean space defined by $x$ whose outward normal derivative on the unit sphere equals $f$ and whose value at the origin equals 0. Here $f$ must be a polynomial function of $x$. Furthermore, the integral of $f$ over the unit sphere with respect to surface area measure must equal 0 (Green's identity shows that this condition is necessary for the existence of a solution to the standard Neumann problem):

*In[104]:=*  `setDimension[x, 3]`

    ··· setDimension:
      x will be considered to be a vector in 3−dimensional real Euclidean space.

*In[105]:=*  `neumann[x_1^6 x_2, x]`

*Out[105]=* $\frac{1}{3003} \left( 143 x_2 - 91 \|x\|^2 x_2 + 33 \|x\|^4 x_2 - 5 \|x\|^6 x_2 + 455 x_1^2 x_2 - 462 \|x\|^2 x_1^2 x_2 + 135 \|x\|^4 x_1^2 x_2 + 693 x_1^4 x_2 - 495 \|x\|^2 x_1^4 x_2 + 429 x_1^6 x_2 \right)$

To check that the last output is the correct solution, we take its Laplacian and its normal derivative, getting the expected results:

*In[106]:=*  `Δ_x[%]`

*Out[106]=* `0`

*In[107]:=*  `normalD[%%, x]`

*Out[107]=* $x_1^6 x_2$

**neumann[f, x]** is computed by using the algorithm discussed in [AR].

**neumann[f, g, x]** is the solution to the generalized Neumann problem: find the function on the unit ball in the Euclidean space defined by $x$ whose outward normal derivative on the unit sphere equals $f$, who Laplacian equals $g$, and whose value at the origin equals 0. Here $f$ and $g$ must be a polynomial functions of $x$. Furthermore, the integral of $f$ over the unit sphere with respect to surface area measure must equal the integral of $g$ over the unit ball with respect to volume measure (Green's identity shows that this condition is necessary for the existence of a solution to the generalized Neumann problem).

In the following example, the necessary condition that the integral of $x_1{}^3 x_2{}^4 x_3{}^2$ over the unit sphere with respect to surface area measure equals the integral of $5 x_1{}^2 x_2{}^3$ over the unit ball with respect to volume measure is satisfied because both integrals equal 0 (by symmetry, because each integrand has as a factor a coordinate of $x$ raised to an odd power):

*In[108]:=*
```
neumann[ x₁³ x₂⁴ x₃², 5 x₁² x₂³, x]
```

*Out[108]=*
$$\frac{1}{45\,945\,900} \big(119\,340\, x_1 - 85\,680\, \|x\|^2\, x_1 + 40\,392\, \|x\|^4\, x_1 - 10\,800\, \|x\|^6\, x_1 + 1260\, \|x\|^8\, x_1 +$$
$$35\,700\, x_1^3 - 47\,124\, \|x\|^2\, x_1^3 + 24\,300\, \|x\|^4\, x_1^3 - 4620\, \|x\|^6\, x_1^3 - 2\,552\,550\, x_2 + 1\,624\,350\, \|x\|^2\, x_2 -$$
$$589\,050\, \|x\|^4\, x_2 + 89\,250\, \|x\|^6\, x_2 - 4\,060\,875\, x_1^2\, x_2 + 4\,123\,350\, \|x\|^2\, x_1^2\, x_2 - 1\,204\,875\, \|x\|^4\, x_1^2\, x_2 +$$
$$214\,200\, x_1\, x_2^2 - 282\,744\, \|x\|^2\, x_1\, x_2^2 + 145\,800\, \|x\|^4\, x_1\, x_2^2 - 27\,720\, \|x\|^6\, x_1\, x_2^2 + 282\,744\, x_1^3\, x_2^2 -$$
$$356\,400\, \|x\|^2\, x_1^3\, x_2^2 + 120\,120\, \|x\|^4\, x_1^3\, x_2^2 - 1\,353\,625\, x_2^3 + 1\,374\,450\, \|x\|^2\, x_2^3 - 401\,625\, \|x\|^4\, x_2^3 -$$
$$12\,370\,050\, x_1^2\, x_2^3 + 8\,835\,750\, \|x\|^2\, x_1^2\, x_2^3 + 141\,372\, x_1\, x_2^4 - 178\,200\, \|x\|^2\, x_1\, x_2^4 +$$
$$60\,060\, \|x\|^4\, x_1\, x_2^4 + 386\,100\, x_1^3\, x_2^4 - 300\,300\, \|x\|^2\, x_1^3\, x_2^4 + 107\,100\, x_1\, x_3^2 - 141\,372\, \|x\|^2\, x_1\, x_3^2 +$$
$$72\,900\, \|x\|^4\, x_1\, x_3^2 - 13\,860\, \|x\|^6\, x_1\, x_3^2 + 141\,372\, x_1^3\, x_3^2 - 178\,200\, \|x\|^2\, x_1^3\, x_3^2 + 60\,060\, \|x\|^4\, x_1^3\, x_3^2 +$$
$$848\,232\, x_1\, x_2^2\, x_3^2 - 1\,069\,200\, \|x\|^2\, x_1\, x_2^2\, x_3^2 + 360\,360\, \|x\|^4\, x_1\, x_2^2\, x_3^2 + 2\,316\,600\, x_1^3\, x_2^2\, x_3^2 -$$
$$1\,801\,800\, \|x\|^2\, x_1^3\, x_2^2\, x_3^2 + 1\,158\,300\, x_1\, x_2^4\, x_3^2 - 900\,900\, \|x\|^2\, x_1\, x_2^4\, x_3^2 + 5\,105\,100\, x_1^3\, x_2^4\, x_3^2\big)$$

To check that the last output is the correct solution, we take its Laplacian and its normal derivative, getting the expected results:

*In[109]:=*
```
Δₓ[%]
```

*Out[109]=*
$$5\, x_1^2\, x_2^3$$

*In[110]:=*   `normalD[%%, x]`

*Out[110]=*   $x_1^3 \, x_2^4 \, x_3^2$

**neumann[f, x, region-> quadratic[b,c,d]]** is the solution to the following Neumann problem: Let $q(x) = b.x^2 + c.x + d$. Find the harmonic function $h$ on the unit ball in the Euclidean space defined by $x$ whose normal derivative on the ellipsoid $\{x \in \mathbf{R}^n : q(x) = 0\}$ equals $\dfrac{f}{\|\nabla q\|}$ and such that $h(0) = 0$. Here $b$ and $c$ should be lists of numbers or symbols of length dimension[$x$], and $d$ should be a number or symbol.

Here $f$ must be a polynomial function of $x$. Furthermore, the integral of $\dfrac{f}{\|\nabla q\|}$ over the ellipsoid $\{x \in \mathbf{R}^n : q(x) = 0\}$ with respect to surface area measure (the function integrateEllipsoidArea can compute this integral) must equal 0 (Green's identity shows that this condition is necessary for the existence of a solution to the standard Neumann problem; see Theorem 2.2 of [AS]).

Because the normal derivative on the ellipsoid $\{x \in \mathbf{R}^n : q(x) = 0\}$ of a function $h$ equals $\dfrac{\nabla h.\nabla q}{\|\nabla q\|}$, this version of the Neumann problem asks to find a harmonic function $h$ such that $\nabla h.\nabla q = f$ and $h(0) = 0$.

Using the notation above, if $c$ consists of all 0's, then it may be omitted. Similarly, if $d = -1$, then $d$ may be omitted (independently of whether or not $c$ has been omitted). For example, we have the following Neumann problem on the ellipsoid $\{x \in \mathbf{R}^3 : 5\,x_1^2 + 3\,x_2^2 + 2\,x_3^2 - 1 = 0\}$:

*In[111]:=*   `neumann[ x₁³ x₂ x₃², x, region -> quadratic[{5, 3, 2}]]`

*Out[111]=*   $\dfrac{1}{144\,767\,520} \big(36\,900\,x_1\,x_2 - 20\,470\,x_1^3\,x_2 - 197\,775\,x_1^5\,x_2 - 103\,410\,x_1\,x_2^3 - 21\,330\,x_1^3\,x_2^3 +$ $84\,321\,x_1\,x_2^5 + 371\,640\,x_1\,x_2\,x_3^2 + 2\,041\,740\,x_1^3\,x_2\,x_3^2 - 779\,220\,x_1\,x_2^3\,x_3^2 - 631\,260\,x_1\,x_2\,x_3^4\big)$

Notice the huge integers appearing in the solution of a problem whose input data contains only single-digit integers.

First we check that our alleged solution above is actually harmonic:

*In[112]:=*
```
Δₓ[%]
```

*Out[112]=*
```
0
```

Now we want to verify that the gradient of our alleged solution dotted with the gradient of $q$ agrees with $x_1{}^3 x_2 x_3{}^2$ on the ellipsoid $\{x \in \mathbf{R}^3 : q(x) = 0\}$:

*In[113]:=*
```
gradient[%%, x].{10 x₁, 6 x₂, 4 x₃}
```

*Out[113]=*
$$\frac{1}{402\,132}\left(1640\,x_1\,x_2 - 2047\,x_1^3\,x_2 - 30\,765\,x_1^5\,x_2 - 8043\,x_1\,x_2^3 - 2844\,x_1^3\,x_2^3 + 9369\,x_1\,x_2^5 + 24\,776\,x_1\,x_2\,x_3^2 + 249\,546\,x_1^3\,x_2\,x_3^2 - 77\,922\,x_1\,x_2^3\,x_3^2 - 56\,112\,x_1\,x_2\,x_3^4\right)$$

The result above does not look much like $x_1{}^3 x_2 x_3{}^2$, so we subtract $x_1{}^3 x_2 x_3{}^2$ to see if we get a function that equals 0 on the ellipsoid $\{x \in \mathbf{R}^3 : 5\,x_1{}^2 + 3\,x_2{}^2 + 2\,x_3{}^2 - 1 = 0\}$:

*In[114]:=*
```
% - x₁³ x₂ x₃²
```

*Out[114]=*
$$\frac{1}{402\,132}\left(1640\,x_1\,x_2 - 2047\,x_1^3\,x_2 - 30\,765\,x_1^5\,x_2 - 8043\,x_1\,x_2^3 - 2844\,x_1^3\,x_2^3 + 9369\,x_1\,x_2^5 + 24\,776\,x_1\,x_2\,x_3^2 - 152\,586\,x_1^3\,x_2\,x_3^2 - 77\,922\,x_1\,x_2^3\,x_3^2 - 56\,112\,x_1\,x_2\,x_3^4\right)$$

The result above does not look much like 0, but it factors nicely:

*In[115]:=*
```
Factor[%]
```

*Out[115]=*
$$-\frac{1}{402\,132}x_1\,x_2\left(-1 + 5\,x_1^2 + 3\,x_2^2 + 2\,x_3^2\right)\left(1640 + 6153\,x_1^2 - 3123\,x_2^2 + 28\,056\,x_3^2\right)$$

The expression above is clearly 0 on the ellipsoid $\{x \in \mathbf{R}^3 : 5\,x_1{}^2 + 3\,x_2{}^2 + 2\,x_3{}^2 - 1 = 0\}$, showing that we indeed have the correct solution.

To illustrate a more complicated example involving an ellipsoid not centered at the origin, let $q(x) = 5\,x_1^2 + 3\,x_2^2 + 2\,x_3^2 + x_1 + 4\,x_2 + 6\,x_3 - 7$. We start with the function $x_1{}^3 x_3$, but this function does not satisfy the necessary integral condition of this ellipsoid for there to exist a solution for the corresponding Neumann problem (see Theorem 2.2 of [AS]). Thus we subtract an appropriate constant k, which we can find as follows:

*In[116]:=* `Solve[ integrateEllipsoidArea[ x₁³ x₃ - k, {5, 3, 2}, {1, 4, 6}, -7, x] == 0, k]`

*Out[116]=* $\left\{\left\{k \to \frac{97}{250}\right\}\right\}$

Now that we know that $\frac{97}{250}$ is the appropriate constant to subtract, we can solve the appropriate Neumann problem on the ellipsoid $\{x \in \mathbf{R}^3 : q(x) = 0\}$:

*In[117]:=* `neumann[ x₁³ x₃ - 97/250, x, region -> quadratic[{5, 3, 2}, {1, 4, 6}, -7]]`

*Out[117]=*
$\frac{1}{14\,968\,128\,000}$
$(-1\,365\,215\,424\,x_1 + 27\,518\,085\,x_1^2 + 61\,268\,550\,x_1^3 - 53\,498\,340\,x_2 + 178\,613\,400\,x_1\,x_2 -$
$\quad 40\,123\,755\,x_2^2 + 133\,960\,050\,x_1\,x_2^2 - 618\,086\,615\,x_3 + 1\,417\,403\,900\,x_1\,x_3 - 81\,779\,250\,x_1^2\,x_3 +$
$\quad 206\,034\,500\,x_1^3\,x_3 + 27\,713\,000\,x_2\,x_3 - 245\,014\,000\,x_1\,x_2\,x_3 + 20\,784\,750\,x_2^2\,x_3 -$
$\quad 183\,760\,500\,x_1\,x_2^2\,x_3 + 12\,605\,670\,x_3^2 - 317\,765\,700\,x_1\,x_3^2 + 20\,331\,500\,x_3^3 - 144\,781\,000\,x_1\,x_3^3)$

Let's check that our alleged solution above is actually harmonic:

*In[118]:=* `Δₓ[%]`

*Out[118]=* `0`

Now we want to verify that the gradient of our alleged solution dotted with the gradient of $q$ agrees with $x_1^3\,x_3$ on the ellipsoid $\{x \in \mathbf{R}^3 : q(x) = 0\}$:

*In[119]:=* `gradient[%%, x].{10 x₁ + 1, 6 x₂ + 4, 4 x₃ + 6}`

*Out[119]=*
$\frac{1}{1\,069\,152\,000}$
$(-377\,694\,891 - 312\,731\,505\,x_1 + 17\,392\,275\,x_1^2 + 219\,590\,250\,x_1^3 - 21\,220\,620\,x_2 + 175\,672\,200\,x_1\,x_2 -$
$\quad 15\,915\,465\,x_2^2 + 131\,754\,150\,x_1\,x_2^2 - 56\,630\,180\,x_3 + 1\,063\,346\,550\,x_1\,x_3 - 96\,042\,750\,x_1^2\,x_3 +$
$\quad 500\,369\,500\,x_1^3\,x_3 + 14\,171\,000\,x_2\,x_3 - 455\,026\,000\,x_1\,x_2\,x_3 + 10\,628\,250\,x_2^2\,x_3 -$
$\quad 341\,269\,500\,x_1\,x_2^2\,x_3 + 10\,646\,190\,x_3^2 - 594\,702\,900\,x_1\,x_3^2 + 7\,085\,500\,x_3^3 - 227\,513\,000\,x_1\,x_3^3)$

The result above does not look much like $x_1^3\,x_2\,x_3^2 - \frac{97}{250}$, so we subtract $x_1^3\,x_2\,x_3^2 - \frac{97}{250}$ to see if we get a function that equals 0 on the ellipsoid $\{x \in \mathbf{R}^3 : q(x) = 0\}$:

*In[120]:=*     `% - ( x₁³ x₃ - 97/250 )`

*Out[120]=*

$$\frac{1}{213\,830\,400}$$
$$\big(7\,427\,217 - 62\,546\,301\,x_1 + 3\,478\,455\,x_1^2 + 43\,918\,050\,x_1^3 - 4\,244\,124\,x_2 + 35\,134\,440\,x_1\,x_2 -$$
$$3\,183\,093\,x_2^2 + 26\,350\,830\,x_1\,x_2^2 - 11\,326\,036\,x_3 + 212\,669\,310\,x_1\,x_3 - 19\,208\,550\,x_1^2\,x_3 -$$
$$113\,756\,500\,x_1^3\,x_3 + 2\,834\,200\,x_2\,x_3 - 91\,005\,200\,x_1\,x_2\,x_3 + 2\,125\,650\,x_2^2\,x_3 -$$
$$68\,253\,900\,x_1\,x_2^2\,x_3 + 2\,129\,238\,x_3^2 - 118\,940\,580\,x_1\,x_3^2 + 1\,417\,100\,x_3^3 - 45\,502\,600\,x_1\,x_3^3\big)$$

The result above does not look much like 0, but it factors nicely:

*In[121]:=*     **Factor[%]**

*Out[121]=*

$$-\frac{1}{213\,830\,400}$$
$$\big(1\,061\,031 - 8\,783\,610\,x_1 - 708\,550\,x_3 + 22\,751\,300\,x_1\,x_3\big)\ \big(-7 + x_1 + 5\,x_1^2 + 4\,x_2 + 3\,x_2^2 + 6\,x_3 + 2\,x_3^2\big)$$

The expression above is clearly 0 on the ellipsoid
$\{x \in \mathbf{R}^3 : 5\,x_1^2 + 3\,x_2^2 + 2\,x_3^2 + x_1 + 4\,x_2 + 6\,x_3 - 7 = 0\}$, showing that we indeed have the correct solution.

**neumann[f, g, x, region-> quadratic[b,c,d]]** is the solution to the following Neumann problem: Let $q(x) = b.x^2 + c.x + d$. Find the harmonic function $h$ on the unit ball in the Euclidean space defined by $x$ whose normal derivative on the ellipsoid $\{x \in \mathbf{R}^n : q(x) = 0\}$ equals $\frac{f}{\|\nabla q\|}$, whose Laplacian equals $g$, and such that $h(0) = 0$. Here $b$ and $c$ should be lists of numbers or symbols of length dimension[$x$], and $d$ should be a number or symbol.

Here $f$ must be a polynomial function of $x$. Furthermore, the integral of $\frac{f}{\|\nabla q\|}$ over the ellipsoid $\{x \in \mathbf{R}^n : q(x) = 0\}$ with respect to surface area measure (the function integrateEllipsoidArea can compute this integral) must equal the integral of $g$ with respect to volume measure over $\{x \in \mathbf{R}^n : q(x) < 0\}$ (Green's identity shows that this condition is necessary for the existence of a solution to the standard Neumann problem; see Theorem 2.9 of [AS]).

Because the normal derivative on the ellipsoid $\{x \in \mathbf{R}^n : q(x) = 0\}$ of a function $h$ equals $\frac{\nabla h.\nabla q}{\|\nabla q\|}$, this version of the generalized Neumann problem asks to find a function $h$ such that $\nabla h.\nabla q = f$, $\Delta h = g$, and $h(0) = 0$.

Using the notation above, if $c$ consists of all 0's, then it may be omitted. Similarly, if $d = -1$, then $d$ may be omitted (independently of whether or not $c$ has been omitted).

In the following example, the necessary condition that the integral of $x_1{}^3 x_2{}^2 x_3$ over $\left\{ x \in \mathbf{R}^3 : 5 x_1{}^2 + 3 x_2{}^2 + 2 x_3{}^2 = 1 \right\}$ with respect to surface area measure equals the integral of $4 x_2{}^3$ over $\left\{ x \in \mathbf{R}^3 : 5 x_1{}^2 + 3 x_2{}^2 + 2 x_3{}^2 < 1 \right\}$ with respect to volume measure is satisfied because both integrals equal 0 (by symmetry, because each integrand has as a factor a coordinate of $x$ raised to an odd power):

*In[122]:=* 
```
neumann[ x₁³ x₂² x₃, 4 x₂³, x, region → quadratic[ {5, 3, 2}]]
```

*Out[122]=* 
$$-\frac{1\,621\,829\,x_2}{34\,123\,734} + \frac{457\,865\,x_1^2\,x_2}{5\,687\,289} - \frac{4825\,x_1^4\,x_2}{66\,518} - \frac{1\,660\,991\,x_2^3}{34\,123\,734} + \frac{34\,955\,x_1^2\,x_2^3}{199\,554} + \frac{17\,278\,x_2^5}{99\,777} + \frac{1\,505\,411\,x_1\,x_3}{5\,687\,742\,816} - \frac{94\,163\,x_1^3\,x_3}{812\,534\,688} - \frac{4355\,x_1^5\,x_3}{3\,224\,344} + \frac{564\,709\,x_1\,x_2^2\,x_3}{270\,844\,896} + \frac{11\,033\,x_1^3\,x_2^2\,x_3}{806\,086} - \frac{16\,629\,x_1\,x_2^4\,x_3}{3\,224\,344} + \frac{745\,261\,x_2\,x_3^2}{11\,374\,578} - \frac{6005\,x_1^2\,x_2\,x_3^2}{66\,518} + \frac{18\,593\,x_2^3\,x_3^2}{199\,554} - \frac{235\,273\,x_1\,x_3^3}{406\,267\,344} - \frac{97\,x_1^3\,x_3^3}{1\,612\,172} - \frac{5437\,x_1\,x_2^2\,x_3^3}{1\,612\,172} - \frac{1049\,x_2\,x_3^4}{33\,259} + \frac{716\,x_1\,x_3^5}{2\,015\,215}$$

To check that the last output is the correct solution, we first take its Laplacian, getting the expected result :

*In[123]:=* 
```
Δₓ[%]
```

*Out[123]=* 
$4 x_2^3$

Now we want to verify that the gradient of our alleged solution dotted with the gradient of $5 x_1{}^2 + 3 x_2{}^2 + 2 x_3{}^2 - 1$ agrees with $x_1{}^3 x_2{}^2 x_3$ on the ellipsoid $\left\{ x \in \mathbf{R}^3 : 5 x_1{}^2 + 3 x_2{}^2 + 2 x_3{}^2 = 1 \right\}$:

*In[124]:=* 
```
gradient[%%, x] . {10 x₁, 6 x₂, 4 x₃}
```

*Out[124]=* 
$$\frac{1}{256\,728\,866\,287\,824} \left( -73\,210\,684\,472\,464\,x_2 + 537\,378\,392\,663\,840\,x_1^2\,x_2 - 856\,624\,851\,507\,600\,x_1^4\,x_2 - 224\,935\,467\,325\,968\,x_2^3 + 1\,708\,862\,692\,812\,240\,x_1^2\,x_2^3 + 1\,333\,702\,562\,230\,080\,x_2^5 + 951\,300\,824\,531\,x_1\,x_3 - 1011\,560\,811\,091\,x_1^3\,x_3 - 18\,724\,716\,557\,820\,x_1^5\,x_3 + 13\,917\,207\,563\,571\,x_1\,x_2^2\,x_3 + 161\,638\,486\,167\,312\,x_1^3\,x_2^2\,x_3 - 50\,313\,330\,111\,492\,x_1\,x_2^4\,x_3 + 235\,491\,827\,710\,832\,x_2\,x_3^2 - 788\,002\,234\,432\,560\,x_1^2\,x_2\,x_3^2 + 621\,923\,665\,189\,008\,x_2^3\,x_3^2 - 3\,270\,826\,887\,526\,x_1\,x_3^3 - 648\,760\,430\,808\,x_1^3\,x_3^3 - 29\,437\,544\,358\,936\,x_1\,x_2^2\,x_3^3 - 178\,140\,917\,531\,808\,x_2\,x_3^4 + 2\,736\,450\,476\,928\,x_1\,x_3^5 \right)$$

The result above does not look much like $x_1{}^3 x_2{}^2 x_3$, so we subtract $x_1{}^3 x_2{}^2 x_3$ to see if we get a function that equals 0 on the ellipsoid $\{x \in \mathbf{R}^3 : 5 x_1{}^2 + 3 x_2{}^2 + 2 x_3{}^2 = 1\}$:

*In[125]:=*
```
% - x₁^3 x₂^2 x₃
```

*Out[125]=*
$$\frac{1}{256\,728\,866\,287\,824} \left( -73\,210\,684\,472\,464\, x_2 + 537\,378\,392\,663\,840\, x_1^2\, x_2 - \right.$$
$$856\,624\,851\,507\,600\, x_1^4\, x_2 - 224\,935\,467\,325\,968\, x_2^3 + 1\,708\,862\,692\,812\,240\, x_1^2\, x_2^3 +$$
$$1\,333\,702\,562\,230\,080\, x_2^5 + 951\,300\,824\,531\, x_1\, x_3 - 1\,011\,560\,811\,091\, x_1^3\, x_3 -$$
$$18\,724\,716\,557\,820\, x_1^5\, x_3 + 13\,917\,207\,563\,571\, x_1\, x_2^2\, x_3 - 95\,090\,380\,120\,512\, x_1^3\, x_2^2\, x_3 -$$
$$50\,313\,330\,111\,492\, x_1\, x_2^4\, x_3 + 235\,491\,827\,710\,832\, x_2\, x_3^2 - 788\,002\,234\,432\,560\, x_1^2\, x_2\, x_3^2 +$$
$$621\,923\,665\,189\,008\, x_2^3\, x_3^2 - 3\,270\,826\,887\,526\, x_1\, x_3^3 - 648\,760\,430\,808\, x_1^3\, x_3^3 -$$
$$\left. 29\,437\,544\,358\,936\, x_1\, x_2^2\, x_3^3 - 178\,140\,917\,531\,808\, x_2\, x_3^4 + 2\,736\,450\,476\,928\, x_1\, x_3^5 \right)$$

The result above does not look much like 0, but it factors nicely:

*In[126]:=*
```
Factor[%]
```

*Out[126]=*
$$\frac{1}{256\,728\,866\,287\,824}$$
$$\left( -1 + 5 x_1^2 + 3 x_2^2 + 2 x_3^2 \right) \left( 73\,210\,684\,472\,464\, x_2 - 171\,324\,970\,301\,520\, x_1^2\, x_2 + 444\,567\,520\,743\,360\, x_2^3 - \right.$$
$$951\,300\,824\,531\, x_1\, x_3 - 3\,744\,943\,311\,564\, x_1^3\, x_3 - 16\,771\,110\,037\,164\, x_1\, x_2^2\, x_3 -$$
$$\left. 89\,070\,458\,765\,904\, x_2\, x_3^2 + 1\,368\,225\,238\,464\, x_1\, x_3^3 \right)$$

The expression above is clearly 0 on the ellipsoid $\{x \in \mathbf{R}^3 : 5 x_1{}^2 + 3 x_2{}^2 + 2 x_3{}^2 = 1\}$, showing that we indeed have the correct solution.

The solutions to these Neumann problems on ellipsoids is computed by using the algorithm developed in [AS]. The *Mathematica* code in the HFT package for the crucial case of ellipsoids of the form $\{x \in \mathbf{R}^n : b.x^2 - 1 = 0\}$ was written by Peter Shin.

## exteriorNeumann

**exteriorNeumann[p, x]** is the solution to the standard exterior Neumann problem: find the harmonic function on the exterior of the unit ball in the Euclidean space defined by $x$ whose outward normal derivative (with respect to the exterior of the unit ball) on the unit sphere equals $p$ and whose limit at $\infty$ equals 0. Here $p$ must be a polynomial function of $x$. If the dimension of the Euclidean space equals 2, then the integral of $p$ over the unit circle with respect to arc length measure must

equal 0:

*In[127]:=*    `setDimension[x, 5]`

     **⋯ setDimension**:

       x will be considered to be a vector in 5-dimensional real Euclidean space.

*In[128]:=*    `exteriorNeumann[x₁^6 x₂, x]`

*Out[128]=*    $\dfrac{5\,x_2}{924\,\|x\|^5} - \dfrac{15\,\left(\|x\|^2\,x_2 - 7\,x_1^2\,x_2\right)}{2002\,\|x\|^9} + \dfrac{\|x\|^4\,x_2 - 18\,\|x\|^2\,x_1^2\,x_2 + 33\,x_1^4\,x_2}{264\,\|x\|^{13}} +$

       $\dfrac{-\frac{1}{143}\,\|x\|^6\,x_2 + \frac{3}{13}\,\|x\|^4\,x_1^2\,x_2 - \|x\|^2\,x_1^4\,x_2 + x_1^6\,x_2}{10\,\|x\|^{17}}$

*In[129]:=*    `Δ_x[%]`

*Out[129]=*    `0`

*In[130]:=*    `normalD[%%, x]`

*Out[130]=*    $-x_1^6\,x_2$

Note that the normal derivative above is the negative of our desired function. This result is correct because the function **`normalD`** takes the outward normal with respect to the unit ball, but for the exterior Neumann problem the outward normal points in the opposite direction.

---

## **biDirichlet**

**`biDirichlet[p, x]`** is the solution to the standard biDirichlet problem: find the biharmonic function on the unit ball in the Euclidean space defined by *x* that equals *p* on the unit sphere and whose normal derivative on the unit sphere equals 0. (A function is called biharmonic if the Laplacian of its Laplacian equals 0.) Here *p* must be a polynomial function of *x*:

In[131]:= **setDimension[x, 3]**

⋯ **setDimension**:

x will be considered to be a vector in 3–dimensional real Euclidean space.

In[132]:= **biDirichlet$\left[x_1^4 x_2^3, x\right]$**

Out[132]= $\frac{1}{30\,030}$ $\left(1287\, x_2 - 4524\, \|x\|^2\, x_2 + 5922\, \|x\|^4\, x_2 - 3420\, \|x\|^6\, x_2 + 735\, \|x\|^8\, x_2 + 13\,650\, x_1^2\, x_2 -\right.$

$40\,530\, \|x\|^2\, x_1^2\, x_2 + 40\,110\, \|x\|^4\, x_1^2\, x_2 - 13\,230\, \|x\|^6\, x_1^2\, x_2 + 24\,255\, x_1^4\, x_2 - 48\,510\, \|x\|^2\, x_1^4\, x_2 +$

$24\,255\, \|x\|^4\, x_1^4\, x_2 + 2275\, x_2^3 - 6755\, \|x\|^2\, x_2^3 + 6685\, \|x\|^4\, x_2^3 - 2205\, \|x\|^6\, x_2^3 +$

$48\,510\, x_1^2\, x_2^3 - 97\,020\, \|x\|^2\, x_1^2\, x_2^3 + 48\,510\, \|x\|^4\, x_1^2\, x_2^3 + 135\,135\, x_1^4\, x_2^3 - 105\,105\, \|x\|^2\, x_1^4\, x_2^3\left.\right)$

To check that the last output is correct, first we take its Laplacian squared to check that we have a biharmonic function:

In[133]:= **$\Delta_x^2$[%]**

Out[133]= **0**

Next we check that the outward normal derivative on the unit sphere indeed equals 0:

In[134]:= **normalD[%%, x]**

Out[134]= **0**

Finally, we check that we have the correct boundary values:

In[135]:= **%%% /. ‖x‖ → 1**

Out[135]= $x_1^4\, x_2^3$

**biDirichlet[p, x]** is computed by using the algorithm discussed in [AR].

# Spherical Harmonics

## basisH

**basisH[m, x]** gives a vector space basis of $H_m(\mathbf{R}^n)$, the space of homogeneous harmonic polynomials on $\mathbf{R}^n$ of degree $m$, where $n$ is the dimension of $x$. The user must first set the dimension of $x$ to an positive integer value. Wrapping the *Mathematica* command **TableForm** around **basisH[m, x]** will produce an output display that is more readable than the usual list format:

*In[136]:=*     **setDimension[x, 3]**

    ••• setDimension:

     x will be considered to be a vector in 3-dimensional real Euclidean space.

*In[137]:=*     **TableForm[basisH[4, x]]**

*Out[137]//TableForm=*

$3 \, \|x\|^4 - 30 \, \|x\|^2 \, x_2^2 + 35 \, x_2^4$
$3 \, \|x\|^2 \, x_2 \, x_3 - 7 \, x_2^3 \, x_3$
$\|x\|^4 - 5 \, \|x\|^2 \, x_2^2 - 5 \, \|x\|^2 \, x_3^2 + 35 \, x_2^2 \, x_3^2$
$3 \, \|x\|^2 \, x_2 \, x_3 - 7 \, x_2 \, x_3^3$
$3 \, \|x\|^4 - 30 \, \|x\|^2 \, x_3^2 + 35 \, x_3^4$
$3 \, \|x\|^2 \, x_1 \, x_2 - 7 \, x_1 \, x_2^3$
$\|x\|^2 \, x_1 \, x_3 - 7 \, x_1 \, x_2^2 \, x_3$
$\|x\|^2 \, x_1 \, x_2 - 7 \, x_1 \, x_2 \, x_3^2$
$3 \, \|x\|^2 \, x_1 \, x_3 - 7 \, x_1 \, x_3^3$

An optional third argument for **basisH** will produce a basis of $H_m(\mathbf{R}^n)$ that is orthonormal with respect to the inner product determined by the third argument. Two inner products are already defined for possible use. One of them will be generated by using **Ball** for the third argument, giving the inner product on $L^2(B, dV)$ (here $B$ denotes the open unit ball in Euclidean space and $V$ denotes volume measure on $B$). Thus **basisH[4, x, Ball]** produces a basis of $H_4(\mathbf{R}^3)$ that is orthonormal with respect to the inner product on $L^2(B, dV)$, where $B$ is the open unit ball in $\mathbf{R}^3$ (the 3 comes from the dimension of $x$, which was set to 3 above), as shown below:

*In[138]:=* `TableForm[basisH[4, x, Ball]]`

*Out[138]//TableForm=*

$$\frac{3}{16}\sqrt{\frac{11}{\pi}}\;\left(3\,\|x\|^4 - 30\,\|x\|^2\,x_2^2 + 35\,x_2^4\right)$$

$$\frac{3}{4}\sqrt{\frac{55}{2\pi}}\;\left(3\,\|x\|^2\,x_2\,x_3 - 7\,x_2^3\,x_3\right)$$

$$\frac{3}{8}\sqrt{\frac{55}{\pi}}\;\left(\|x\|^4 - 8\,\|x\|^2\,x_2^2 + 7\,x_2^4 - 2\,\|x\|^2\,x_3^2 + 14\,x_2^2\,x_3^2\right)$$

$$\frac{3}{4}\sqrt{\frac{385}{2\pi}}\;\left(3\,\|x\|^2\,x_2\,x_3 - 3\,x_2^3\,x_3 - 4\,x_2\,x_3^3\right)$$

$$\frac{3}{16}\sqrt{\frac{385}{\pi}}\;\left(\|x\|^4 - 2\,\|x\|^2\,x_2^2 + x_2^4 - 8\,\|x\|^2\,x_3^2 + 8\,x_2^2\,x_3^2 + 8\,x_3^4\right)$$

$$\frac{3}{4}\sqrt{\frac{55}{2\pi}}\;\left(3\,\|x\|^2\,x_1\,x_2 - 7\,x_1\,x_2^3\right)$$

$$\frac{3}{4}\sqrt{\frac{55}{\pi}}\;\left(\|x\|^2\,x_1\,x_3 - 7\,x_1\,x_2^2\,x_3\right)$$

$$\frac{3}{4}\sqrt{\frac{385}{2\pi}}\;\left(\|x\|^2\,x_1\,x_2 - x_1\,x_2^3 - 4\,x_1\,x_2\,x_3^2\right)$$

$$\frac{3}{4}\sqrt{\frac{385}{\pi}}\;\left(\|x\|^2\,x_1\,x_3 - x_1\,x_2^2\,x_3 - 2\,x_1\,x_3^3\right)$$

The other inner product that is already defined is generated by using **Sphere** as the third argument, giving the inner product on $L^2(S,\ d\sigma)$, where $S$ denotes the unit sphere in Euclidean space and $\sigma$ denotes normalized surface area measure on $S$. Thus **basisH[4, x, Sphere]** produces a basis of $H_4(\mathbf{R}^3)$ that is orthonormal with respect to the inner product on $L^2(S, d\sigma)$, where $S$ is the open unit sphere in $\mathbf{R}^3$ (the 3 comes from the dimension of $x$, which was set to 3 above), as shown below:

*In[139]:=* `TableForm[basisH[4, x, Sphere]]`

*Out[139]//TableForm=*

$\frac{3}{8} \left( 3 \|x\|^4 - 30 \|x\|^2 x_2^2 + 35 x_2^4 \right)$

$\frac{3}{2} \sqrt{\frac{5}{2}} \left( 3 \|x\|^2 x_2 x_3 - 7 x_2^3 x_3 \right)$

$\frac{3}{4} \sqrt{5} \left( \|x\|^4 - 8 \|x\|^2 x_2^2 + 7 x_2^4 - 2 \|x\|^2 x_3^2 + 14 x_2^2 x_3^2 \right)$

$\frac{3}{2} \sqrt{\frac{35}{2}} \left( 3 \|x\|^2 x_2 x_3 - 3 x_2^3 x_3 - 4 x_2 x_3^3 \right)$

$\frac{3}{8} \sqrt{35} \left( \|x\|^4 - 2 \|x\|^2 x_2^2 + x_2^4 - 8 \|x\|^2 x_3^2 + 8 x_2^2 x_3^2 + 8 x_3^4 \right)$

$\frac{3}{2} \sqrt{\frac{5}{2}} \left( 3 \|x\|^2 x_1 x_2 - 7 x_1 x_2^3 \right)$

$\frac{3}{2} \sqrt{5} \left( \|x\|^2 x_1 x_3 - 7 x_1 x_2^2 x_3 \right)$

$\frac{3}{2} \sqrt{\frac{35}{2}} \left( \|x\|^2 x_1 x_2 - x_1 x_2^3 - 4 x_1 x_2 x_3^2 \right)$

$\frac{3}{2} \sqrt{35} \left( \|x\|^2 x_1 x_3 - x_1 x_2^2 x_3 - 2 x_1 x_3^3 \right)$

More generally, the optional third argument for **basisH** can be any inner product defined on polynomials on $\mathbf{R}^n$. This inner product should itself be a function of three arguments, giving the inner product of the first two arguments; the third argument specifies the variable. For example, the following can be used to find an orthonormal basis in the weighted space $L^2 \left( B, \ \left( 1 - \|x\|^2 \right) dV \right)$.

*In[140]:=* `innerProduct[f_, g_, x_] := integrateBall[ f g (1 - ||x||^2), x]`

*In[141]:=* `TableForm[basisH[4, x, innerProduct]]`

*Out[141]//TableForm=*

$$\frac{3}{16} \sqrt{\frac{143}{2\pi}} \ \left(3 \|x\|^4 - 30 \|x\|^2 x_2^2 + 35 x_2^4\right)$$

$$\frac{3}{8} \sqrt{\frac{715}{\pi}} \ \left(3 \|x\|^2 x_2 x_3 - 7 x_2^3 x_3\right)$$

$$\frac{3}{8} \sqrt{\frac{715}{2\pi}} \ \left(\|x\|^4 - 8 \|x\|^2 x_2^2 + 7 x_2^4 - 2 \|x\|^2 x_3^2 + 14 x_2^2 x_3^2\right)$$

$$\frac{3}{8} \sqrt{\frac{5005}{\pi}} \ \left(3 \|x\|^2 x_2 x_3 - 3 x_2^3 x_3 - 4 x_2 x_3^3\right)$$

$$\frac{3}{16} \sqrt{\frac{5005}{2\pi}} \ \left(\|x\|^4 - 2 \|x\|^2 x_2^2 + x_2^4 - 8 \|x\|^2 x_3^2 + 8 x_2^2 x_3^2 + 8 x_3^4\right)$$

$$\frac{3}{8} \sqrt{\frac{715}{\pi}} \ \left(3 \|x\|^2 x_1 x_2 - 7 x_1 x_2^3\right)$$

$$\frac{3}{4} \sqrt{\frac{715}{2\pi}} \ \left(\|x\|^2 x_1 x_3 - 7 x_1 x_2^2 x_3\right)$$

$$\frac{3}{8} \sqrt{\frac{5005}{\pi}} \ \left(\|x\|^2 x_1 x_2 - x_1 x_2^3 - 4 x_1 x_2 x_3^2\right)$$

$$\frac{3}{4} \sqrt{\frac{5005}{2\pi}} \ \left(\|x\|^2 x_1 x_3 - x_1 x_2^2 x_3 - 2 x_1 x_3^3\right)$$

To produce an orthonormal basis of $H_m(S)$, the space of spherical harmonics on the unit sphere $S$ with respect to the inner product on $L^2(S, \ d\sigma)$, use the optional third argument **Sphere** and in the result replace $\|x\|$ with 1:

*In[142]:=* `TableForm[basisH[4, x, Sphere]] /. ‖x‖ → 1`

*Out[142]//TableForm=*

$$\frac{3}{8} \left(3 - 30 x_2^2 + 35 x_2^4\right)$$

$$\frac{3}{2} \sqrt{\frac{5}{2}} \left(3 x_2 x_3 - 7 x_2^3 x_3\right)$$

$$\frac{3}{4} \sqrt{5} \left(1 - 8 x_2^2 + 7 x_2^4 - 2 x_3^2 + 14 x_2^2 x_3^2\right)$$

$$\frac{3}{2} \sqrt{\frac{35}{2}} \left(3 x_2 x_3 - 3 x_2^3 x_3 - 4 x_2 x_3^3\right)$$

$$\frac{3}{8} \sqrt{35} \left(1 - 2 x_2^2 + x_2^4 - 8 x_3^2 + 8 x_2^2 x_3^2 + 8 x_3^4\right)$$

$$\frac{3}{2} \sqrt{\frac{5}{2}} \left(3 x_1 x_2 - 7 x_1 x_2^3\right)$$

$$\frac{3}{2} \sqrt{5} \left(x_1 x_3 - 7 x_1 x_2^2 x_3\right)$$

$$\frac{3}{2} \sqrt{\frac{35}{2}} \left(x_1 x_2 - x_1 x_2^3 - 4 x_1 x_2 x_3^2\right)$$

$$\frac{3}{2} \sqrt{35} \left(x_1 x_3 - x_1 x_2^2 x_3 - 2 x_1 x_3^3\right)$$

**basisH** is computed by using Theorem 5.25 of [ABR]. If the optional third argument is present in **basisH**, then the Gram-Schmidt procedure is used to produce an orthonormal basis with respect to the inner product determined by the third argument.

---

## zonalHarmonic

**zonalHarmonic[m, x, y]** is the extended zonal harmonic $Z_m(x, y)$ as defined by 8.7 of [ABR]:

*In[143]:=* `setDimension[x, n]`

⋯ setDimension:

x will be considered to be a vector in n−dimensional real Euclidean space.

*In[144]:=*    `zonalHarmonic[5, x, y]`

*Out[144]=*    $\frac{1}{120}$ n (2 + n) (4 + n) (6 + n) (8 + n) (x.y)$^5$ −

$\frac{1}{12}$ n (2 + n) (4 + n) (8 + n) (x.y)$^3$ $\|x\|^2$ $\|y\|^2$ + $\frac{1}{8}$ n (2 + n) (8 + n) x.y $\|x\|^4$ $\|y\|^4$

To obtain the zonal harmonic $Z_m(x, y)$ as a homogeneous polynomial in $x$ of degree $m$ with pole $z$ on the unit sphere (as in Theorem 5.38 of [ABR]), use **`zonalHarmonic[m, x, y]`** and in the result replace $\|z\|$ with 1:

*In[145]:=*    `setDimension[x, 7]`

     ⋯ **setDimension**:

     x will be considered to be a vector in 7−dimensional real Euclidean space.

*In[146]:=*    `zonalHarmonic[6, x, z] /. ‖z‖ → 1`

*Out[146]=*    $\frac{357}{16}$ $\left(143 \ (x.z)^6 - 143 \ (x.z)^4 \ \|x\|^2 + 33 \ (x.z)^2 \ \|x\|^4 - \|x\|^6\right)$

To obtain the zonal harmonic $Z_m(x, z)$ as an element of $H_m(S)$ as originally defined in Chapter 5 of [ABR], use **`zonalHarmonic[m,x,z]`** and in the result replace $\|x\|$ and $\|z\|$ with 1:

*In[147]:=*    `zonalHarmonic[8, x, z] /. {‖x‖ → 1, ‖z‖ → 1}`

*Out[147]=*    $\frac{693}{128}$ $\left(7 - 364 \ (x.z)^2 + 2730 \ (x.z)^4 - 6188 \ (x.z)^6 + 4199 \ (x.z)^8\right)$

---

## dimHarmonic

**`dimHarmonic[m, n]`** is the vector space dimension of $H_m(\mathbf{R}^n)$, the space of harmonic polynomials on $\mathbf{R}^n$ homogeneous of degree $m$:

*In[148]:=*  `dimHarmonic[12, 100]`

*Out[148]=*  3 901 030 682 812 965

**dimHarmonic** is computed by using Proposition 5.8 of [ABR].

---

# Inversion and the Kelvin Transform

---

## reflection

**reflection[x]** is the reflection of a vector $x$ with respect to the unit sphere. Thus the reflection of $x$ equals $\frac{x}{\|x\|^2}$. The argument of **reflection** can be either an expression representing a vector or a list of coordinates for a vector:

*In[149]:=*  `reflection[2 x + y]`

*Out[149]=*  $\dfrac{2\,x + y}{\|2\,x + y\|^2}$

*In[150]:=*  `reflection[{1, -2, 5, 11}]`

*Out[150]=*  $\left\{\dfrac{1}{151}, -\dfrac{2}{151}, \dfrac{5}{151}, \dfrac{11}{151}\right\}$

**reflection[x, Sphere[c, r]]** is the reflection of a vector $x$ with respect to the sphere centered at $c$ with radius $r$. Here each of **x** and **c** can be either an expression representing a vector or a list of coordinates for a vector:

*In[151]:=* `reflection[{2, 4, 5}, Sphere[{3, 1, 6}, 7]]`

*Out[151]=* $\left\{-\frac{16}{11}, \frac{158}{11}, \frac{17}{11}\right\}$

*In[152]:=* `reflection[2 x + y, Sphere[z, 3]]`

*Out[152]=* $z + \frac{9 (2 x + y - z)}{\|2 x + y - z\|^2}$

**`reflection[x, Sphere[c, r]]`** is computed using equation 4.13 in [ABR].

**`reflection[x, Hyperplane[b, t]]`** is the reflection of a vector $x$ with respect to the hyperplane consisting of those vectors whose inner product with $b$ equals $t$. In other words, if $n$ denotes the dimension of $x$ (which must equal the dimension of $b$), then the reflection is taken with respect to the hyperplane $\{y \in \mathbf{R}^n : b.y = t\}$. Here each of **`x`** and **`b`** can be either an expression representing a vector or a list of coordinates for a vector:

*In[153]:=* `reflection[x, Hyperplane[{1, 4, 5}, 7]]`

*Out[153]=* $\left\{\frac{1}{21} (7 + 20 x_1 - 4 x_2 - 5 x_3), \frac{1}{21} (28 - 4 x_1 + 5 x_2 - 20 x_3), \frac{1}{21} (35 - 5 x_1 - 20 x_2 - 4 x_3)\right\}$

*In[154]:=* `reflection[{x_1, x_2}, Hyperplane[{4, 5}, 7]]`

*Out[154]=* $\left\{\frac{1}{41} (56 + 9 x_1 - 40 x_2), \frac{1}{41} (70 - 40 x_1 - 9 x_2)\right\}$

**`reflection[x, Hyperplane[b, t]]`** is computed using Exercise 9 on page 71 of [ABR].

# kelvin

**`kelvin[u, x]`** is the Kelvin transform of $u$, thought of as a function of $x$:

*In[155]:=*
```
setDimension[x, n]
```

··· setDimension:
x will be considered to be a vector in n−dimensional real Euclidean space.

*In[156]:=*
```
kelvin[ (x.y)² + x₃⁵ ‖x‖, x]
```

*Out[156]=*
$\|\mathbf{x}\|^{-9-n} \left( (\mathbf{x}.\mathbf{y})^2 \|\mathbf{x}\|^7 + \mathbf{x}_3^5 \right)$

# Phi

$\Phi[\mathbf{z}]$ is is the modified inversion introduced on page 154 of [ABR]. The modified inversion $\Phi$ is defined by

$$\Phi(z) = 2\, \frac{z - \mathbf{southPole}}{\|z - \mathbf{southPole}\|^2} + \mathbf{southPole},$$

where **southPole** is the south pole $(0, 0, ..., 0, -1)$. The modified inversion $\Phi$ is useful for translating questions about harmonic functions on balls to questions about harmonic functions on half-spaces, and vice versa.

Here is a proof of that $\Phi(\Phi(z)) = z$, as claimed in Proposition 7.18 (a) of [ABR]:

*In[157]:=*
```
Φ[z]
```

··· setDimension: southPole will be considered to
be a vector in dimension[z]−dimensional real Euclidean space.

*Out[157]=*
$\mathbf{southPole} + \dfrac{2\,(-\mathbf{southPole} + \mathbf{z})}{\|-\mathbf{southPole} + \mathbf{z}\|^2}$

*In[158]:=*
```
Φ[%]
```

*Out[158]=*
```
z
```

Another, and often more convenient, form of $\Phi$ is given by $\Phi[\mathbf{x}, \mathbf{y}]$. Here $x$ is a vector, $y$ is a real number, and thus $(x, y)$ is a vector in a Euclidean space whose dimension is one more than the dimension of $x$.

Here is a proof of the assertion of Exercise 6 in Chapter 7 of [ABR], which asserts that

$$1 - \|\Phi(x, y)\|^2 = \frac{4\,y}{\|x\|^2 + (y + 1)^2}:$$

*In[159]:=*  `Φ[x, y]`

*Out[159]=*  $\left\{\dfrac{2\,x}{(1 + y)^2 + \|x\|^2},\ -1 + \dfrac{2\,(1 + y)}{(1 + y)^2 + \|x\|^2}\right\}$

*In[160]:=*  `1 - ‖%‖²`

*Out[160]=*  $\dfrac{4\,y}{1 + 2\,y + y^2 + \|x\|^2}$

## kelvinH

**kelvinH[u, z]** is the modified Kelvin transform defined to be equal to

$$2^{(n-2)/2}\,\|z - \textbf{southPole}\|^{2-n}\,u(\Phi(z)).$$

This modified Kelvin transform is defined on page 155 of [ABR]. Here is a proof of the assertion made there that the modified Kelvin transform is its own inverse:

*In[161]:=*  `setDimension[z, n]`

⋯ setDimension:
  z will be considered to be a vector in n−dimensional real Euclidean space.

*In[162]:=*  `kelvinH[u[z], z]`

*Out[162]=*  $2^{-1 + \frac{n}{2}}\,\|-\text{southPole} + z\|^{2-n}\,u\!\left[\text{southPole} + \dfrac{2\,(-\text{southPole} + z)}{\|-\text{southPole} + z\|^2}\right]$

*In[163]:=*   `kelvinH[%, z]`

*Out[163]=*   $\left(\dfrac{1}{\|-southPole + z\|}\right)^{-n} \|-southPole + z\|^{-n} u[z]$

*In[164]:=*   `PowerExpand[%]`

*Out[164]=*   $u[z]$

Another form of `kelvinH` is given by `kelvinH[u, x, y]`. Here $x$ is a vector, $y$ is a real variable, and $u$ is a function of $(x, y)$, which is a vector in a Euclidean space whose dimension is one more than the dimension of $x$:

*In[165]:=*   `setDimension[x, n - 1]`

    ••• setDimension:

      x will be considered to be a vector in $-1 + n$–dimensional real Euclidean space.

*In[166]:=*   `kelvinH[z[4], z]`

*Out[166]=*   $2^{-1+\frac{n}{2}} \|-southPole + z\|^{-n} \left(-2\, southPole_4 + \|-southPole + z\|^2\, southPole_4 + 2\, z_4\right)$

The output above includes terms of the form `southPole`$_4$, which denotes the fourth coordinate of `southPole`. Of course `southPole`$_4$ equals either 0 or -1, but since the dimension of $z$ has been set to equal the symbol $n$, the computer has no way of determining whether or not $n$ equals 4. With the input `kelvinH[x[4], x, y]` there is no ambiguity; $x_4$ cannot be the last coordinate, because the last coordinate is denoted by $y$, and thus the fourth coordinate of `southPole` is 0:

*In[167]:=*   `kelvinH[x[4], x, y ]`

*Out[167]=*   $2^{n/2} \left((1 + y)^2 + \|x\|^2\right)^{-n/2} x_4$

# Kernels

## **poissonKernel**

**poissonKernel[x, y]** is the extended Poisson kernel $P(x,y)$ for the unit ball as defined by Equation 6.21 of [ABR]:

*In[168]:=*    **setDimension[x, n]**

     ⋯   setDimension:

     x will be considered to be a vector in n−dimensional real Euclidean space.

*In[169]:=*    **poissonKernel[x, y]**

*Out[169]=*    $\left(1 - \|x\|^2 \|y\|^2\right) \left(1 - 2\,x.y + \|x\|^2 \|y\|^2\right)^{-n/2}$

The usual Poisson kernel $P(x, \zeta)$, where $\zeta$ lies on the unit sphere, can be obtained by replacing $\|\zeta\|$ with 1:

*In[170]:=*    **poissonKernel[x, ζ] /. ‖ζ‖ → 1**

*Out[170]=*    $\left(1 - \|x\|^2\right) \left(1 - 2\,x.\zeta + \|x\|^2\right)^{-n/2}$

## **poissonKernelH**

**poissonKernelH[z, w]** is the extended Poisson kernel $P_H(z, w)$ for the upper half-space $H$, as defined by Equation 8.22 of [ABR]:

*In[171]:=*    **setDimension[z, 5]**

     ⋯   setDimension:

     z will be considered to be a vector in 5−dimensional real Euclidean space.

*In[172]:=* `poissonKernelH[z, w]`

*Out[172]=* $\dfrac{3\,(w_5 + z_5)}{4\,\pi^2\,\left(-2\,w.z + \|w\|^2 + \|z\|^2 + 4\,w_5\,z_5\right)^{5/2}}$

The usual Poisson kernel $P_H(z, w)$ for the upper half-space, where we think of $w \in \partial H$, can be obtained by replacing the last coordinate of $w$ with 0:.

*In[173]:=* `% /. w₅ → 0`

*Out[173]=* $\dfrac{3\,z_5}{4\,\pi^2\,\left(-2\,w.z + \|w\|^2 + \|z\|^2\right)^{5/2}}$

**`poissonKernelH[x, y, t, u]`** equals $P_H\big((x, y), (t, u)\big)$ and thus is another format for the extended Poisson kernel on the upper half-space. Here $x$ and $t$ denote vectors, and $y$ and $u$ denote nonnegative numbers. Note that if we want to work in the upper half-space of $\mathbf{R}^n$, then the dimension of $x$ should be set to $n - 1$:

*In[174]:=* `setDimension[x, n - 1]`

    •••   setDimension:

      x will be considered to be a vector in $-1 + n$–dimensional real Euclidean space.

*In[175]:=* `poissonKernelH[x, y, t, u]`

*Out[175]=* $\dfrac{2\,(u + y)\,\left((u + y)^2 + \|-t + x\|^2\right)^{-n/2}}{n \, \text{volume}[n]}$

The usual Poisson kernel $P_H\big((x, y), t\big)$ for the upper half-space, where $t \in \partial H$, can be setting the last argument in this format to 0:

*In[176]:=* `poissonKernelH[x, y, t, 0]`

*Out[176]=* $\dfrac{2\,y\,\left(y^2 + \|-t + x\|^2\right)^{-n/2}}{n \, \text{volume}[n]}$

## **bergmanKernel**

**bergmanKernel[x, y]** is the reproducing kernel for the harmonic Bergman space of the unit ball:

*In[177]:=*
```
setDimension[x, 10]
```

••• setDimension:
x will be considered to be a vector in 10−dimensional real Euclidean space.

*In[178]:=*
```
bergmanKernel[x, y]
```

*Out[178]=*
$$\frac{12 \left(10 + 8 \, (-3 + x.y) \, \|x\|^2 \, \|y\|^2 + 6 \, \|x\|^4 \, \|y\|^4\right)}{\pi^5 \left(1 - 2 \, x.y + \|x\|^2 \, \|y\|^2\right)^6}$$

**bergmanKernel** is computed using Theorem 8.13 of [ABR].

## **bergmanKernelH**

**bergmanKernelH[z, w]** is the reproducing kernel for the harmonic Bergman space of the upper half-space:

*In[179]:=*
```
setDimension[z, 10]
```

••• setDimension:
z will be considered to be a vector in 10−dimensional real Euclidean space.

*In[180]:=*
```
bergmanKernelH[z, w]
```

*Out[180]=*
$$\frac{48 \left(2 \, w.z - \|w\|^2 - \|z\|^2 + 16 \, w_{10} \, z_{10} + 10 \left(w_{10}^2 + z_{10}^2\right)\right)}{\pi^5 \left(2 \, w.z - \|w\|^2 - \|z\|^2 - 4 \, w_{10} \, z_{10}\right)^6}$$

**bergmanKernelH[x, y, t, u]** is another format for the reproducing kernel for the harmonic Bergman space of the upper half-space. Here *x* and *t* denote vectors, *y* and *u* denote nonnegative numbers, and the kernel is evaluated at $((x, y), (t, u))$. Note that if we want to work in the upper half-space of $\mathbf{R}^n$, then the dimension of $x$ should be set to $n - 1$:

*In[181]:=* `setDimension[x, n - 1]`

   ••• setDimension:
     x will be considered to be a vector in $-1 + n$–dimensional real Euclidean space.

*In[182]:=* `bergmanKernelH[x, y, t, u]`

*Out[182]=* $\frac{1}{n \, \text{volume}[n]} 4 \left( (-1 + n) \, (u + y)^2 - \|-t + x\|^2 \right) \left( (u + y)^2 + \|-t + x\|^2 \right)^{-1 - \frac{n}{2}}$

**bergmanKernelH** is computed by using Theorem 8.24 of [ABR].

# Miscellaneous

## bergmanProjection

**bergmanProjection[u, x]** is the orthogonal projection of $u$ onto $b^2(B)$, the harmonic Bergman space of the unit ball (see Chapter 8 of [ABR]). Here $u$ must be a polynomial function of $x$:

*In[183]:=* `setDimension[x, 5]`

   ••• setDimension:
     x will be considered to be a vector in 5–dimensional real Euclidean space.

*In[184]:=* `bergmanProjection[x_1^5 x_2^3, x]`

*Out[184]=* $\frac{3 \, x_1 \, x_2}{143} - \frac{3}{221} \, \|x\|^6 \, x_1 \, x_2 + \frac{2}{17} \, \|x\|^4 \, x_1^3 \, x_2 - \frac{3}{17} \, \|x\|^2 \, x_1^5 \, x_2 +$
$\frac{1}{17} \, \|x\|^4 \, x_1 \, x_2^3 - \frac{10}{17} \, \|x\|^2 \, x_1^3 \, x_2^3 + x_1^5 \, x_2^3 + \frac{1}{17} \left( -\|x\|^2 \, x_1 \, x_2 + 2 \, x_1^3 \, x_2 + x_1 \, x_2^3 \right) +$
$\frac{1}{2717} \left( 135 \, \|x\|^4 \, x_1 \, x_2 - 660 \, \|x\|^2 \, x_1^3 \, x_2 + 429 \, x_1^5 \, x_2 - 330 \, \|x\|^2 \, x_1 \, x_2^3 + 1430 \, x_1^3 \, x_2^3 \right)$

**bergmanProjection** is computed by using Corollary 8.15 of [ABR].

---

## harmonicConjugate

**harmonicConjugate[u, x, y]** is the harmonic conjugate of $u$ on $\mathbf{R}^2$, where the coordinates in $\mathbf{R}^2$ are denoted by $x$, $y$ and $u$ is a harmonic function of $(x, y)$:

*In[185]:=*    `harmonicConjugate[15 x^2 y + 12 x^3 y - 5 y^3 - 12 x y^3, x, y]`

*Out[185]=*    $-5 x^3 - 3 x^4 + 15 x y^2 + 18 x^2 y^2 - 3 y^4$

**harmonicConjugate** is computed by using Exercise 11 in Chapter 1 of [ABR].

---

## schwarz

**[x]** is the maximum of $|u(x)|$, where $u$ ranges over all harmonic functions on the unit ball (in the Euclidean space whose dimension is determined by $x$) with $u(0) = 0$ and $|u| < 1$ on the ball. Thus, in the notation used in Theorem 6.24 of [ABR] to present the Harmonic Schwarz Lemma, **schwarz[x]** equals $U(\|x\|\ \mathbf{N})$. The user must first set the dimension of the Euclidean space to a positive integer value:

*In[186]:=*    `setDimension[x, 5]`

       **···**  setDimension:

         x will be considered to be a vector in 5-dimensional real Euclidean space.

*In[187]:=*    `schwarz[x]`

*Out[187]=*    $\left(2 + 4\,\|x\|^2 + 2\,\|x\|^4 - 2\sqrt{1 + \|x\|^2} - 3\,\|x\|^2\sqrt{1 + \|x\|^2} + 3\,\|x\|^4\sqrt{1 + \|x\|^2} + 2\,\|x\|^6\sqrt{1 + \|x\|^2}\right) \Big/ \left(2\,\|x\|^3\,(1 + \|x\|^2)^2\right)$

**schwarz[x]** is computed by evaluating the Poisson integral that defines $U(\|x\|\ \mathbf{N})$, iterating an appropriate formula from Appendix A of [ABR]. The computation of `schwarz[x]` may take a while (this computation involves a complicated integration), so be patient.

## togetherness

When not using the HFT11.m software package, *Mathematica* might tell the user that the result of a calculation is $t^2 + (1 - t)(1 + t)$. Of course the user will easily recognize that the last quantity equals 1, but with more complicated results the user will need to apply appropriate *Mathematica* commands (such as `Simplify`, `Expand`, or `Together`) to get the result in simplest form. Applying `Simplify` to all *Mathematica* output is poor practice, because *Mathematica* can take a long time to realize that some expressions cannot be simplified. For almost all output generated by the HFT11.m package, applying `Together` to the result and all its subexpressions puts it in simplest form. Thus the HFT11.m package redefines the *Mathematica* variable `$Post` so that for each output $X$, *Mathematica* will display either $X$ or the result of applying `Together` to $X$ and all its subexpressions, whichever is shorter (if the user had defined `$Post` before starting the HFT11.m package, the HFT11.m version of `$Post` is composed with the user's `$Post`). This procedure is invisible to the user and usually requires only a short amount of computer time. If you want to turn off this feature, use the command `turnOff[togetherness]` (if the user had defined `$Post` before starting the HFT11.m package, this restores the user's version of `$Post`). The command `turnOn[togetherness]` turns this feature back on again.

If `homogeneous` or `taylor` is used to find an expansion about a point other than the origin, then `togetherness` will be automatically turned off. If this were not done then, for example, the last output in the *Taylor* subsection earlier in this document would be simplified to $1 + x_1 x_2 + x_1{}^2$, which is not what the user wanted to see when asking for an expansion about $b$.

## x, y, z, ... Instead of $x_1$, $x_2$, $x_3$, ...

Most of the examples in this document use symbolic vectors, which we have usually called *x* or *z*. Vectors can also be described by giving explicit coordinates in the form of a list. Both formats (symbolic vectors and explicit lists) work with all functions in this package. Some examples of using explicit coordinates can be found in the subsection earlier in this document titled *Dirichlet problems* *with explicit coordinates*. Here is another example, using $x$, $y$, and $z$ instead of $x_1$, $x_2$, and $x_3$ to find a basis of the space of harmonic polynomials on $\mathbf{R}^3$ that are homogeneous of degree 4:

*In[188]:=*
```
TableForm[basisH[4, {x, y, z}]]
```

*Out[188]//TableForm=*

$35\,y^4 - 30\,y^2\,\left(x^2 + y^2 + z^2\right) + 3\,\left(x^2 + y^2 + z^2\right)^2$
$3\,x^2\,y\,z - 4\,y^3\,z + 3\,y\,z^3$
$x^4 - 3\,x^2\,y^2 - 4\,y^4 - 3\,x^2\,z^2 + 27\,y^2\,z^2 - 4\,z^4$
$3\,x^2\,y\,z + 3\,y^3\,z - 4\,y\,z^3$
$35\,z^4 - 30\,z^2\,\left(x^2 + y^2 + z^2\right) + 3\,\left(x^2 + y^2 + z^2\right)^2$
$3\,x^3\,y - 4\,x\,y^3 + 3\,x\,y\,z^2$
$x^3\,z - 6\,x\,y^2\,z + x\,z^3$
$x^3\,y + x\,y^3 - 6\,x\,y\,z^2$
$3\,x^3\,z + 3\,x\,y^2\,z - 4\,x\,z^3$

The next command verifies that all the polynomials above are indeed harmonic.

*In[189]:=*
```
Table[Δ_{x,y,z}[%[[j]]], {j, Length[%]}]
```

*Out[189]=*
```
{0, 0, 0, 0, 0, 0, 0, 0, 0}
```

Again, using explicit coordinate lists instead of symbolic vectors will work with all functions in this package. Each notation has its advantages, depending upon the context. When working in dimensions bigger than 3 or 4, it is usually easier to use symbolic vectors instead of explicit lists of coordinates.

## **zeroToZero**

The command **turnOn[ zeroToZero ]** coverts a numeric output of 0 to an exact output of 0 with no decimal point. For example, *Mathematica* usually gives the result of the following calculation as 0 with a decimal point:

*In[190]:=*
```
Integrate[E^-x^2, {x, 0, ∞}] - 0.5 √π
```

*Out[190]=*
```
0.
```

The following command changes this behavior:

*In[191]:=*
```
turnOn[ zeroToZero]
```

    •••   zeroToZero: zeroToZero has been turned
        on.  The command  turnOff[zeroToZero]  will turn it back off.

Now we do the same integral again, this time getting 0 with no decimal point:

*In[192]:=*
```
Integrate[E^-x^2, {x, 0, ∞}] - 0.5 √π
```

*Out[192]=*
```
0
```

The command **turnOff[ zeroToZero ]** restores the original behavior of *Mathematica*:

*In[193]:=*
```
turnOff[ zeroToZero ]
```

    •••   zeroToZero: zeroToZero has been turned
        off.  The command  turnOn[zeroToZero]  will turn it back on.

---

# Changes from Earlier Versions

### Changes for HFT11.m, version 11.00

---

# antiLaplacian

The `multiple → quadratic` option for `antiLaplacian` finds an antiLaplacian that is a polynomial multiple of a quadratic expression.

---

# dirichlet

The `region → quadratic` option for `dirichlet` now defaults the constant term in the quadratic to -1 if no constant term is provided.

## integrateBall

`integrateBall[f, x]` now works when the dimension of `x` is either a concrete number (such as `3`) or a symbol (such as `n`).

## integrateEllipsoidArea

`integrateEllipsoidArea` is a new function added for version 11.00. This function computes the integral with respect to surface area measure of $\frac{p(x)}{\|(\nabla q)\,(x)\|}$ over an ellipsoid $\{x \in \mathbf{R}^n : q(x) = 0\}$.

## integrateEllipsoidVolume

`integrateEllipsoidVolume` is a new function added for version 11.00. This function computes the integral with respect to volume measure over an ellipsoid.

## kelvin

When the variable for the function `kelvin` is given as a list (as opposed to a symbol), the functions `Together` and `PowerExpand` have been applied to the result so that powers of the norm of the list will be properly collected and simplified.

## neumann

`neumann` can now compute solutions to generalized Neumann problems in addition to standard Neumann problems.

`neumann` can now compute solutions to a class of Neumann problems on ellipsoids.

## norm

`norm[v_List]` is now defined as $\sqrt{\texttt{Plus @@ v}^2}$ without `Abs[v]`², which was causing differentiation problems in some cases.

## normalD

`normalD` can now compute normal derivatives with respect to an arbitrary surface (not just spheres) defined by an equation of the form $q(z) = c$ for some constant $c$.

### Changes for HFT10.m, version 10.00

The naming scheme for functions introduced by this package is that function names are generally spelled out in full and begin with a lower-case letter, as in **laplacian**. This is a major change from HFT9 and previous versions of the package, where function names began with an upper-case letter. This change was made because new versions of *Mathematica* have introduced new functions beginning with an upper-case letter, which have clashed with names already used by the HFT package. With this change (meaning that HFT functions now begin with a lower-case letter), this issue has disappeared.

Functions in the HFT10 package with a name formed from more than one word begin with a lower-case letter but then have the first letter of additional words begin with an upper-case letter, as in **surfaceArea**.

## annulus, Annulus

The previous option **Annulus** has been replaced by **annulus**.

## antiLaplacian, AntiLaplacian

The previous function **AntiLaplacian** has been replaced by **antiLaplacian**. The ability of **antiLaplacian** to detect a singularity at 0 (and thus automatically to use the option **singularity -> 0** without requiring the user to include this option) has been improved.

## basisH, BasisH

The previous function **BasisH** has been replaced by **basisH**. The syntax for finding a basis that is orthonormal with respect to volume measure on the ball or normalized surface area measure on the sphere has been changed. Furthermore, a new feature has been added: now a basis can be found that is orthonormal with respect to an arbitrary user-specified inner product.

## bergmanKernel, BergmanKernel

The previous function **BergmanKernel** has been replaced by **bergmanKernel**.

## bergmanKernelH, BergmanKernelH

The previous function **BergmanKernelH** has been replaced by **bergmanKernelH**.

## bergmanProjection, BergmanProjection

The previous function **BergmanProjection** has been replaced by **bergmanProjection**.

## biDirichlet, BiDirichlet

The previous function **BiDirichlet** has been replaced by **biDirichlet**.

## dimension, Dimension

The previous function **Dimension** has been replaced by **dimension**.

## dimHarmonic, DimensionH

The previous function **DimensionH** has been replaced by **dimHarmonic**.

## dirichlet, Dirichlet

The previous function **Dirichlet** has been replaced by **dirichlet**.

## divergence, Divergence

The previous function **Divergence** has been replaced by **divergence**.

## expandNorm, ExpandNorm

The previous function **ExpandNorm** has been replaced by **expandNorm**.

## exteriorNeumann, exteriorNeumann

The previous function **ExteriorNeumann** has been replaced by **exteriorNeumann**.

## exteriorSphere, ExteriorSphere

The previous option **ExteriorSphere** has been replaced by **exteriorSphere**.

## gradient, Grad

The previous function **Grad** has been replaced by **gradient**. The function **Grad** now is a built-in *Mathematica* function with a different meaning.

## harmonicConjugate, HarmonicConjugate

The previous function **HarmonicConjugate** has been replaced by **harmonicConjugate**.

## hilbertSchmidt, HilbertSchmidt

The previous function **HilbertSchmidt** has been replaced by **hilbertSchmidt**. Also, some additional functionality has been added. For example, **hilbertSchmidt[ 4 IdentityMatrix[n] ]** now evaluates to $4n$.

## homogeneous, Homogeneous

The previous function **Homogeneous** has been replaced by **homogeneous**.

## IdMatrix

The previous function **IdMatrix** has been deleted. Instead, use the built-in *Mathematica* function

**IdentityMatrix**.

---

## integrateBall, IntegrateBall

The previous function **IntegrateBall** has been replaced by **integrateBall**.

---

## integrateSphere, IntegrateSphere

The previous function **IntegrateSphere** has been replaced by **integrateSphere**.

---

## jacobian, J

The previous function **J** has been replaced by **jacobian**.

---

## kelvin, Kelvin

The previous function **Kelvin** has been replaced by **kelvin**.

---

## kelvinH, KelvinM

The previous function **KelvinM** has been replaced by **kelvinH**.

---

## laplacian, Laplacian

The previous function **Laplacian** has been replaced by **laplacian**. The function **Laplacian** now is a built-in *Mathematica* function with a different meaning.

## multiple, Multiple

The previous option **Multiple** has been replaced by **multiple**.

## neumann, Neumann

The previous function **Neumann** has been replaced by **neumann**.

## norm, Norm

The previous function **Norm** has been replaced by **norm**. The function **Norm** now is a built-in *Mathematica* function with a different meaning.

## normalD, NormalD

The previous function **NormalD** has been replaced by **normalD**.

## NumericQ, NumberQ

All previous uses of **NumberQ** have been replaced by **NumericQ** for greater generality.

## orthonormal, Orthonormal

The previous option **Orthonormal** has been replaced by **orthonormal**.

## partial, Partial

The previous function **Partial** has been replaced by **partial**.

## poissonKernel, PoissonKernel

The previous function **PoissonKernel** has been replaced by **poissonKernel**.

## poissonKernelH, PoissonKernelH

The previous function **PoissonKernelH** has been replaced by **poissonKernelH**.

## quadratic, Quadratic

The previous option value **Quadratic** has been replaced by **quadratic**.

## reflection, Reflection

The previous function **Reflection** has been replaced by **reflection**. Furthermore, for reflections in a sphere, the syntax has been changed to agree with the use of **Sphere[c, r]** by *Mathematica*, so that the first argument now denotes the center of the sphere and the second argument now denotes the radius of the sphere.

## region, Region

The previous option **Region** has been replaced by **region**.

## schwarz, Schwarz

The previous function **Schwarz** has been replaced by **schwarz**.

## setDimension, SetDimension

The previous function **SetDimension** has been replaced by **setDimension**.

## singularity, Singularity

The previous function **Singularity** has been replaced by **singularity**.

## southPole, S

The previous symbol **S** has been replaced by **southPole**.

## surfaceArea, SurfaceArea

The previous function **SurfaceArea** has been replaced by **surfaceArea**.

## taylor, Taylor

The previous function **Taylor** has been replaced by **taylor**.

## `trace, Tr`

The previous function `Tr` has been replaced by `trace`. The function `Tr` now is a built-in *Mathematica* function with a slightly different meaning.

## `togetherness, Togetherness`

The previous function `Togetherness` has been replaced by `togetherness`.

## `turnOff, TurnOff`

The previous function `TurnOff` has been replaced by `turnOff`.

## `turnOn, TurnOn`

The previous function `TurnOn` has been replaced by `turnOn`.

## `volume, Volume`

The previous function `Volume` has been replaced by `volume`. The function `Volume` now is a built-in *Mathematica* function with a different meaning.

## `zeroToZero, ZeroToZero`

The previous function `ZeroToZero` has been replaced by `zeroToZero`.

## zonalHarmonic, ZonalHarmonic

The previous function **ZonalHarmonic** has been replaced by **zonalHarmonic**.

***Changes for HFT9.m, version 9.00***

## Grad

Grad is a built-in function in *Mathematica* 9. Thus it has been unprotected, removed, and then redefined to restore previous functionality.

## Laplacian

Laplacian is a built-in function in *Mathematica* 9. Thus it has been unprotected, removed, and then redefined to restore previous functionality.

***Changes for HFT7.m, version 7.01***

## △ (Laplacian)

$\triangle_{\texttt{x\_Symbol}}$ [x_ .y_] is now correctly coded.

***Changes for HFT7.m, version 7.00***

## δ (delta)

The function **Delta** from earlier versions has been replaced by **δ** to agree better with the usual mathematical notation. Furthermore, the syntax has changed, with **Delta[j]** replaced by $\delta_j$.

See the subsection of this document titled "**gradient**" for an example of the use of $\delta$.

## identityMatrix

**identityMatrix[n]** denotes the $n$-by-$n$ identity matrix.

Both **trace[identityMatrix[n]]** and
**hilbertSchmidt[identityMatrix[n]]** will evaluate as n.

See the subsection of this document titled "**jacobian**" for another example of the use of
**identityMatrix**.

## Δ (Laplacian)

The function **Laplacian** from earlier versions has been replaced by Δ to agree better with the usual mathematical notation. Furthermore, the syntax has changed, with **Laplacian[f, x]** replaced by $\Delta_x$**[f]**.

See the subsection of this document titled "Δ (Laplacian)" for more details.

## ZeroToZero

By default, **ZeroToZero** is now off rather than on when HFTm.7 starts. To turn this feature on, use the command **TurnOn[ZeroToZero]**.

See the subsection of this document titled "ZeroToZero" for more details.

Sheldon Axler
Mathematics Department
San Francisco State University
San Francisco, CA 94132, USA

email: axler@sfsu.edu

website: www.axler.net